**On the origin of quasi-periodicities in the atmospheres of Earth, Jupiter, Saturn and the Sun.**

Ian R. Edmonds

(Physics Department, Queensland University of Technology, Australia (Retired))

iredmonds@aapt.net.au

**Abstract.** This paper explores a possible linkage between solar motion about the solar system barycentre and the quasi-periodicity evident in the pressure and temperature of planet atmospheres. We establish that dominant mid frequency range periodicity in planet atmospheres corresponds closely to the harmonic series $39.5/n = T_A/n$ years where $n = 2,3,4,$ …. We establish that the period $T_A = 39.5$ years is the interval between acceleration impulses experienced by the Sun as it passes close to the barycentre and that the time sequence of impulses generates the spectral harmonic series $T_A/n$ that is observed in the periodicity of climate indices like the NAO, PDO, SOI, and QBO. We develop a model of a simple harmonic oscillator responding to periodic acceleration impulses and show that the response duplicates several features of the QBO. We conclude that oscillatory phenomena observed in solar activity and in planet atmosphere variability could be due to the response of the various natural oscillatory modes to impulsive Sun acceleration associated with planetary motion.

**Highlights**

- The periods of significant components in planet atmosphere variability occur at harmonics of the period 39.5 years.
- Impulsive acceleration of the Sun towards the barycentre occurs at intervals of 39.5 years.
- A model of a simple harmonic oscillator responding to periodic acceleration impulses duplicates anomalous features of the QBO.
- The internal modes of solar system atmospheres are entrained into quasi-periodic oscillations at periods close to the harmonics of 39.5 years.

**1. Introduction**

The fluid surfaces of Earth, Jupiter and Saturn exhibit periodic oscillations of pressure and temperature referred to here as surface oscillations. Surface oscillations are measured in specific regions of a planet's surface so we have, for example on Earth, the Pacific Decadal Oscillation (PDO) that measures the difference in surface water temperature between the north-east and tropical Pacific Ocean, Mantua and Hare (2002), while the North Atlantic Oscillation (NAO) measures the surface sea-level pressure difference between the Sub-tropical (Azores) High and the Sub-polar Low. The NAO affects temperature and precipitation patterns in North America and Western Europe while the warm phase of the PDO causes heavy rains in the Eastern Pacific Ocean and droughts in Asia and Australia, (Trenberth and Jones 2007, Norel et al 2021). The periodicity of surface oscillations has been studied extensively, (Le Mouel et al 2019, 2020, Braganza et al 2009, Moore et al 2006, Meehl et al 2008, Weijer et al 2013, van Loon and Meehl 2014, Tsonis, 2018, Pokrovsky and Pokrovsky 2021). Mitchell (1976) suggested surface variability resulted from, (a), internal stochastic mechanisms that are essentially probabilistic and, (b), external forcing mechanisms that involve some form of resonance between internal modes of the system and external forcing of repetitive of cyclic character. In the latter category diurnal, annual, lunar and solar forcing is included. Von der Heyt et al (2021) provided an updated review of climate variability.

The main focus of this paper is on periodicity of surface oscillations in the 2 to 20 year range. The reason for this is that surface oscillation records on Earth are at the most 200 years long while surface oscillation records on other planets are only about 40 years long, (Antuñano et al 2021, Blake et al 2023). As a result the periods of oscillations longer than 20 years in the records are difficult to assess accurately. The NAO record, one of the longest, extends back to 1824 while the PDO record extends back to 1854. Figure 1 shows the periodograms determined for the NAO and PDO using Fast Fourier Transform (FFT) with padding (adding zeros to the record) extended to 2048 months to allow better location of the long period spectral peaks.

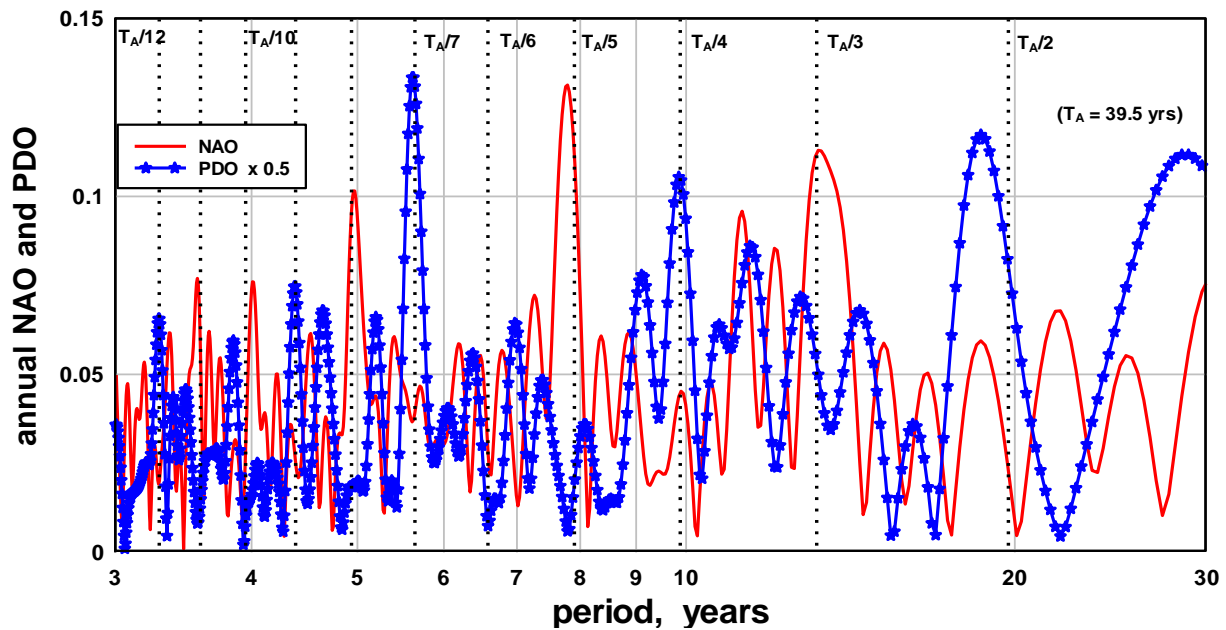

**Figure 1. The spectra of the North Atlantic Oscillation (NAO) and the Pacific Decadal Oscillation (PDO) in the period range 3 to 30 years. The reference lines indicate the periods of the harmonic series $T_A/n = 39.5/n$ years where n = 2, 3, 4…. 12.**

The strongest peaks in the NAO and PDO periodograms coincide closely with harmonics of the period $T_A$ = 39.5 years. The harmonics are indicated in Figure 1 by vertical reference lines labeled $T_A/n$ where n = 2, 3, 4, …. 12. Ten of the eleven harmonic periods shown as reference lines in Figure 1 coincide closely with the strong or moderately strong peaks in the NAO or the PDO spectrums. The exception is the harmonic at $T_A/6$ = 6.58 years where deep minima in the PDO and NAO spectra occur. The absence of a component at $T_A/6$ is significant and is discussed in Section 4. A characteristic of some of the strongest peaks in Figure 1, for example peaks at $T_A/5$ in the NAO and at $T_A/7$ in the PDO, is that they are relatively narrow indicating these periodicities persist over all or most of the record.

The close coincidence of surface oscillation periods with periods given by $T_A/n$ shown in Figure 1 may be just random coincidence. However, Braganza et al (2009) found the two strongest periodicities in the Southern Oscillation Index (SOI) occurred at 3.5 years and 2.9 years, within 3% respectively of $T_A/11$, 3.59 years, and $T_A/14$, 2.82 years. The 28 month, 2.32 year, period Quasi-Biennial Oscillation (QBO) corresponds closely in period to $T_A/17$ = 2.32 years. Pan et al (2020) found "that all the significant peak periods of driving-force signals in climate indices can be represented by harmonics of four base periods: 2.32, 3.90, 6.55, and 11.02 years". Pan et al (2020) attributed the 11.02 year periodicity to the sunspot cycle and assumed the other three periods were linked to intrinsic variability. However, the 2.32, 3.90, and 6.55 year periods are all within 0.5 percent of the $T_A/17$, $T_A/10$, and $T_A/6$ periods respectively suggesting a linkage of climate variability to a source of period $T_A$ = 39.5 years. This remarkable correspondence of climate periodicity with the harmonic series $T_A/n$ could also be construed as random coincidence that appears significant due to teleconnections between the various regions on Earth and the associated oscillations. However, recent observations of the temperature variation of the equatorial atmosphere of Jupiter indicate that Jupiter's Quasi-Quadrennial Oscillation (QQO) is characterized by oscillations at 5.7-year period and 3.9-year period, Antunano et al (2021). The observed periods very closely coincide, respectively, with the $T_A/7$ = 5.64 year harmonic and the $T_A/10$ = 3.95 year harmonic. This suggests that the close coincidence of surface oscillation periodicity with harmonics of $T_A$ may apply to all planet atmospheres. The alignment of climate periodicity with the periods of a simple harmonic series favors deterministic rather than stochastic variability.

The first objective of this paper is to examine if planetary motion is the origin of the period $T_A$ = 39.5 years and if planetary motion could generate a spectrum of periodicities of period $T_A/n$ where n = 2, 3, 4 … A second objective is to examine if the atmospheres of the Sun, Earth, Jupiter and Saturn have dominant periodicities consistent with periods within the spectrum given by $T = T_A/n$ years. A third objective is to develop a model of a single climate oscillation forced by the effects of planetary motion and to compare the model response with the time variation and spectral content of the Earth QBO. An Appendix discusses the possibility of predicting future climate variation by applying the results of this paper.

Section 2 of the paper describes the data sources and methods. Section 3 outlines an elementary theory of planet motion and outlines a scenario by which planet motion could interact with the atmosphere of the Sun and the planet atmospheres to cause oscillations at periods $T_A/n$. Section 4 examines the periodicity of the Group Sunspot Number for evidence of periodicity of the solar surface at $T_A/n$. Section 5 examines the periodicity of the SOI, the QBO, the AO and AAO, the IOD, and Jupiter's QQO for planet surface periodicity at $T_A/n$. Section 6 derives the response of a model oscillator to periodic impulses of acceleration and compares the response with the observed QBO variation. Section 7 is a summary of the results. Section 8 is a discussion and conclusion.

**2. Data sources and methods**

Data on the NAO was obtained at https://www.cpc.ncep.noaa.gov/products/precip/CWlink/pna/norm.nao.monthly.b5001.current.ascii.table

Data on the PDO was obtained at https://www.ncei.noaa.gov/pub/data/cmb/ersst/v5/index/ersst.v5.pdo.dat

Data on the SOI was obtained at http://www.bom.gov.au/climate/enso/soi_monthly.txt

Data on the QBO was obtained at https://psl.noaa.gov/tmp/gcos_wgsp/data.193.116.231.166.176.23.51.30.txt

Data on Group Sunspot number was obtained at http://www.leif.org/research/Preliminary-Revised-GSN-and-B.xls

A wide range of other climate data is available at the NOAA Working Group on Surface Pressure site https://psl.noaa.gov/gcos_wgsp/Timeseries

The methods used for data analysis were as follows. The Fast Fourier Transform as implemented in the DPlot application was used to find the spectral content of the records. The Inverted Notch Filter (INF) was used when narrow band pass filtering was required to examine the time variation of a climate variable in a specified period range. The INF filter method applies a Press digital notch filter to a record, (Press 1986), then subtracts the filtered record from the original record to obtain a narrow band filtered version of the record at a specified frequency and frequency band width. This paper uses the Press digital notch filter as implemented in the DPlot application. The INF method is especially useful for studying quasi-periodic, non-stationary signals.

**3. Elementary theory of planet motion and the associated periodicities.**

As the planets orbit the Sun the distance between the Sun and the centre of mass of the planetary system varies. The centre of mass of the solar system, the barycentre, remains at rest relative to the stars so the Sun moves relative to the barycentre to balance the motion of the planetary centre of mass, Jose (1965). In this article, the centre of mass of the planetary system is calculated from a simplified model of planetary motion where all the planets move about the Sun in circular orbits in the ecliptic plane, Edmonds (2022). This is a good approximation for the planets other than Mercury, the orbit of which is quite elliptical. The variation in centre of mass of the planets, calculated with circular orbits, Edmonds (2022), is scarcely distinguishable from the variation of the centre of mass calculated by taking into account planet eccentricity and inclination (Jose 1965, Cionco 2008, Cionco and Soon 2015).

The time variation of the coordinates, $(x_i, y_i)$, of the ith planet relative to the Sun as origin at (0,0) is calculated using

$$x_i = r_i \cos(\omega_i t + \phi_i), \qquad y_i = r_i \sin(\omega_i t + \phi_i), \qquad (1)$$

where $r_i$ is the orbital radius of the ith planet, the angular frequency $\omega_i = 2\pi/T_i$, the phase angle in radians, $\phi_i = (\pi/180)L_i$, and $L_i$ is the heliographic inertial longitude of the planet in degrees on January 01, 1965. The coordinates, $(x_{PCM}, y_{PCM})$, of the planetary centre of mass (PCM) relative to the Sun are given by

$$x_{PCM} = \Sigma m_i x_i / \Sigma m_i, \quad y_{PCM} = \Sigma m_i y_i / \Sigma m_i \qquad (2)$$

where $m_i$ is the mass of the ith planet. The distance between the Sun and the planetary centre of mass, $r_{PCM}$, is given by

$$r_{PCM} = (x_{PCM}^2 + y_{PCM}^2)^{1/2} \tag{3}$$

The distance between the Sun and the solar system centre of mass or the barycentre is given by

$$R_B = r_{PCM}[\Sigma m_i/\{\Sigma m_i + m_{SUN}\}] \tag{4}$$

where $m_{SUN}$ is the mass of the Sun. The Sun to barycentre distance is usually expressed as the ratio $R_B/R_{SUN}$ where $R_{SUN}$ is the radius of the Sun, 0.0046 AU. Figure 2 shows the time variation of $R_B/R_{SUN}$ for the time interval of interest in this article.

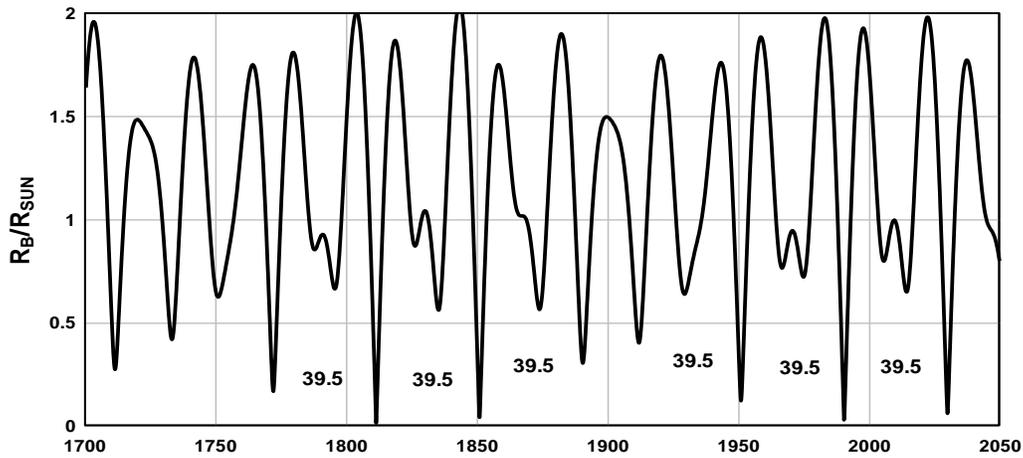

Figure 2. The distance, $R_B$, of the centre of the Sun from the solar system barycentre expressed as a ratio to the radius of the Sun, $R_{SUN}$. The Sun passes close to the barycentre at intervals of 39.5 years e.g. 1950.5 – 1990.

As evident in Figure 2 there is a broad pattern of Sun motion that returns every ~178 years, (Jose 1965, Charvatova 2000). There are times when the centre of the Sun passes very close to the barycentre and $R_B/R_{SUN}$ ~ 0, for example in 1811, 1850, 1950, 1990 and 2030. The time between successive events of close approach is 39.5 years. The spectral content of the $R_B/R_{SUN}$ variation is shown in Figure 3 where, as a means of identifying the contributions of different planets, $R_B/R_{SUN}$ is calculated with the full complement of planets and also with some planets eliminated. The lowest frequency peak, ~ 0.005 years$^{-1}$, is due to conjunctions of Uranus and Neptune; i.e. when Uranus, Neptune and the Sun lie on the same line. The peak at 44.4 years is due to conjunctions of Jupiter, Saturn and Neptune, the peak at 35.7 years due to Saturn and Neptune conjunction and the peak at 20 years due to conjunctions of Jupiter and Saturn. The higher frequency peaks in the spectrum that have been labeled occur close to harmonics of $T_A$ = 39.5 years, for example the peak at 5.61 years is close to $T_A/7$, 5.64 years. An alternative is to view all the peaks in Figure 3 as stable resonances of the planet system, Scafetta et al (2018). In Figure 3 the period $T_A$ = 39.5 years is not represented by a peak but occurs at a deep minimum between the 44.4 year peak and the 35.7 year peak. The $T_A$ period is marked by the dashed reference line.

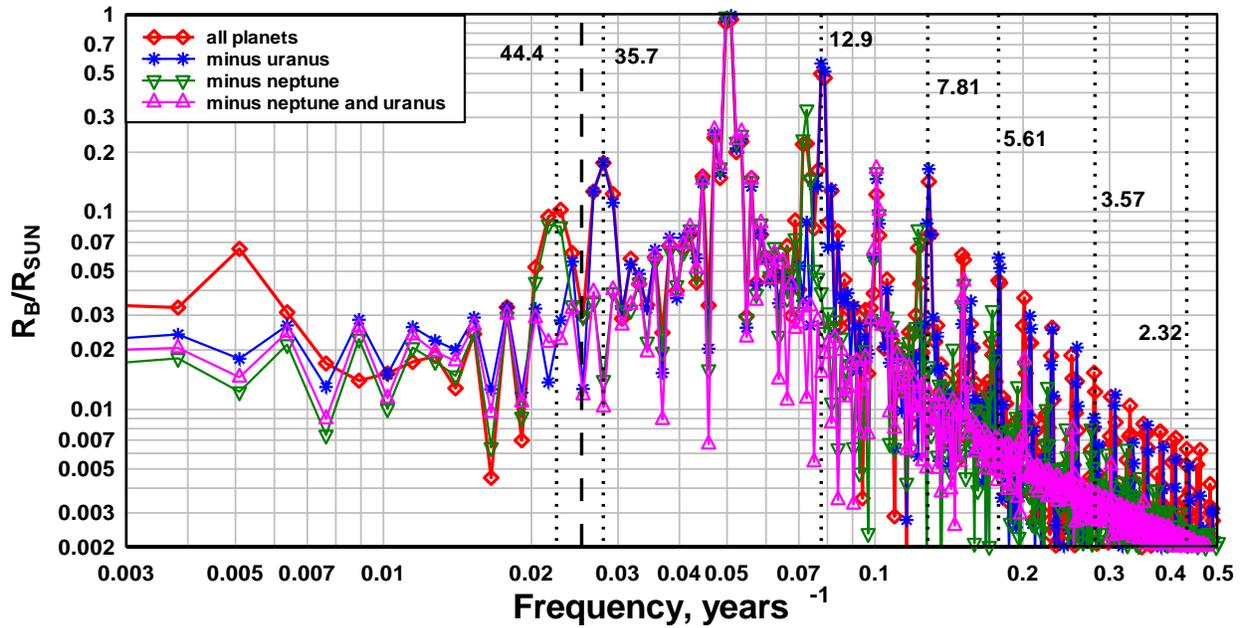

Figure 3. The spectrum of $R_B/R_{SUN}$ calculated for different combinations of outer planets allows the contribution of Uranus and Neptune to the spectral content to be assessed. Periods of some prominent peaks are indicated with reference lines.

Further insight about the spectra of Sun motion is obtained by comparing the spectra of the displacement, speed and acceleration of the Sun as it moves around the barycentre, Figure 4. Increasing measures of rate of change of motion shift the balance of power in the spectral components from low to high frequencies and from the greater effect of the outer planets to the greater effect of the inner planets. For example, in Figure 4 the strongest peaks in Sun acceleration around the barycentre occur at periods of 1.09 years and 0.645 years, 235 days.

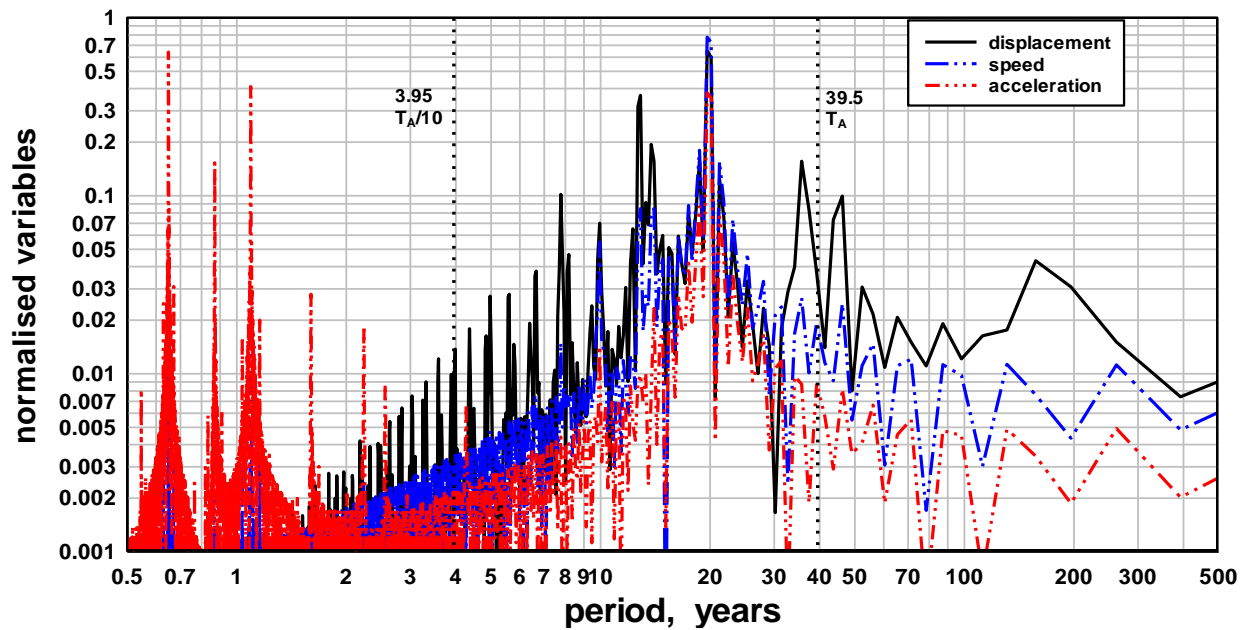

Figure 4. The three periodograms show the variation of normalized displacement, speed and acceleration of the Sun as it moves around the solar system barycentre. The shift from long to short period content reflects the greater contribution of the fast moving inner planets to the speed and acceleration of the Sun.

An alternative view of spectral content of Sun motion is obtained by calculating the acceleration of the Sun along the Sun to barycentre direction. This can be calculated by double differentiating the $R_B/R_{SUN}$ variation in Figure 2. The result for the acceleration of the Sun along the Sun to barycentre direction is shown in Figure 5 and the resulting periodogram of the acceleration is shown in Figure 6. The peaks in the time variation of acceleration are sharp and, as a consequence, the peak heights of the spectral content in Figure 6 are much more even in amplitude as compared with peak heights in Figure 2 and Figure 3 where, as the period increases from 1 to 10 years, the amplitudes of the $T_A/n$ harmonics increase by two orders of magnitude.

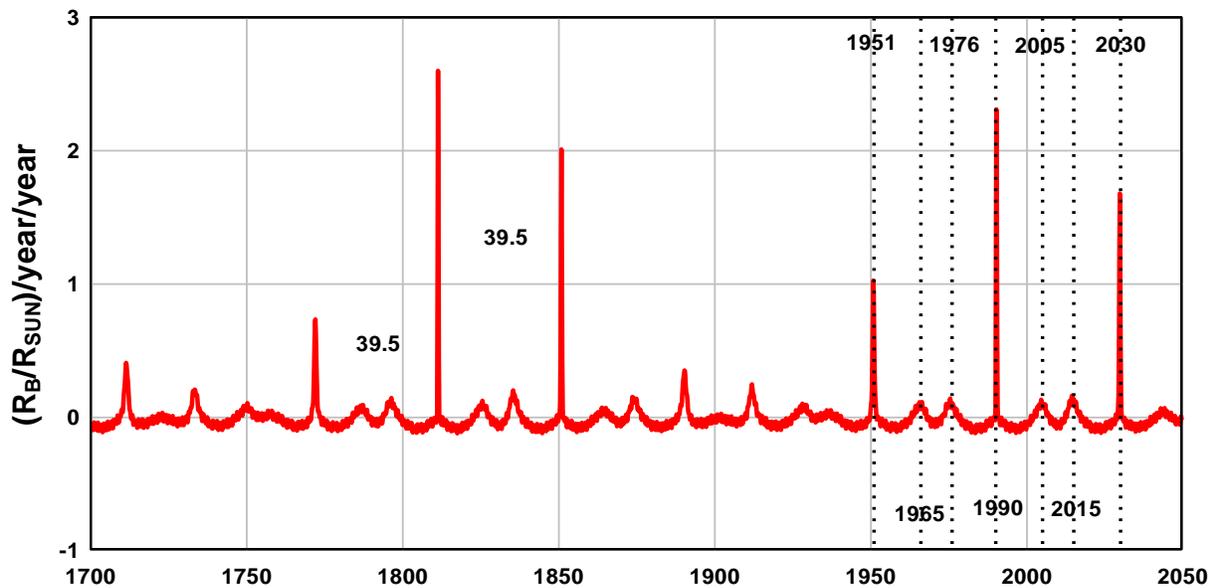

Figure 5. The acceleration of the Sun along the direction towards the barycentre peaks sharply at times when the Sun moves close to the barycentre. The spacing in time between consecutive sharp peaks is 39.5 years.

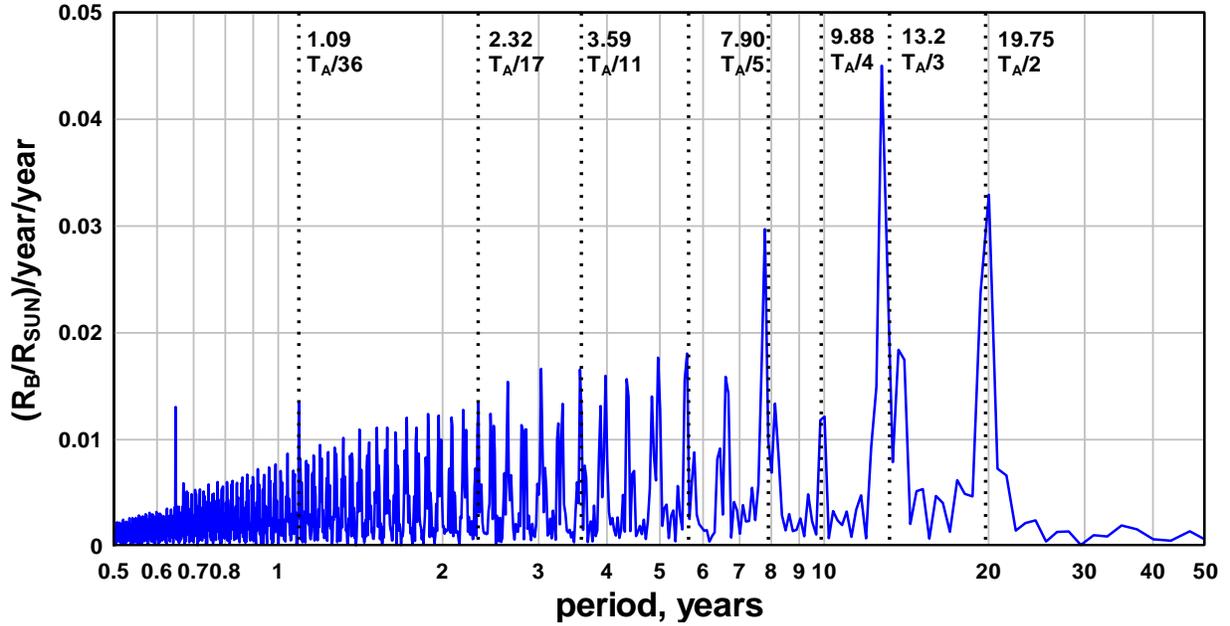

**Figure 6.** The spectrum of Sun acceleration along the Sun – barycentre direction. The extreme sharpness of the acceleration spikes in Figure 5 leads to a spectral content in Figure 6 that is relatively much more even in peak amplitude than the spectral content of acceleration of the Sun around the barycentre, Figure 4.

The time variation of Sun acceleration as shown in Figure 5 is quite complicated and it is not clear how different parts of the variation contribute to the spectrum. Understanding how the spectrum arises is assisted by simulating just the basic features of the time variation in Sun- barycentre acceleration. The simulation in Figure 7 is calculated from a function designed to produce very sharp peaks at spacing 39.5 years, $[1 +\cos(2\pi t/39.5)]^m$ where $m = 12$. The resulting peaks are adjusted so that the central peak of each group of three peaks is separated by 179 years. The periodogram of this simulation is shown in Figure 8 for comparison with the actual harmonics of Sun acceleration towards the barycentre, Figure 6. The periods, $T_A/n$, are shown as reference lines in Figure 8.

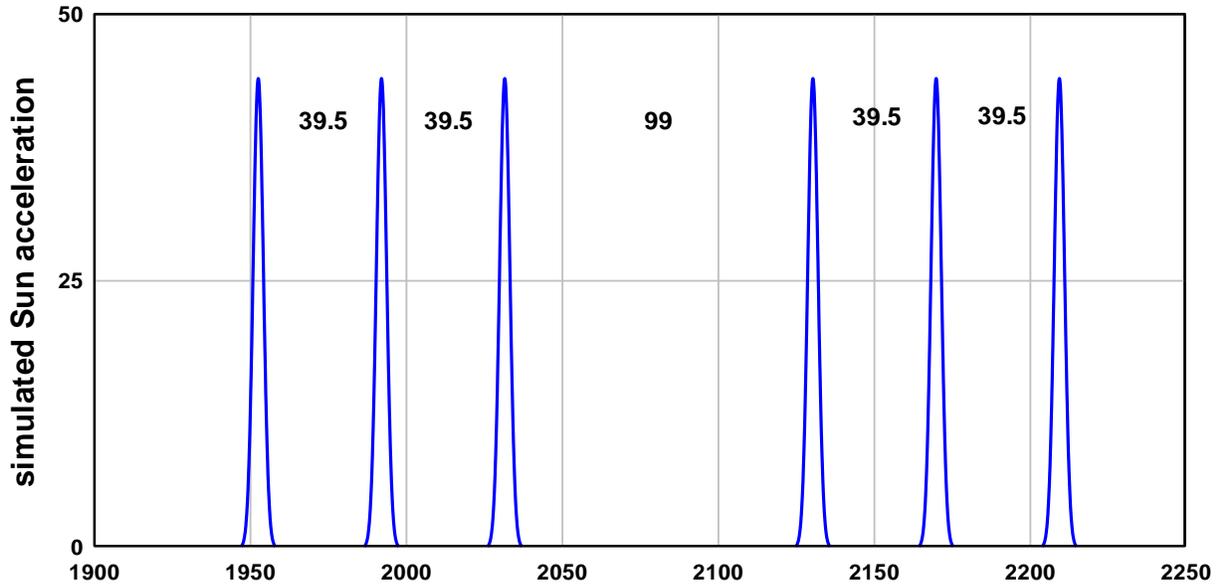

**Figure 7.** Simulation of the acceleration of the Sun in the Sun – barycentre direction.

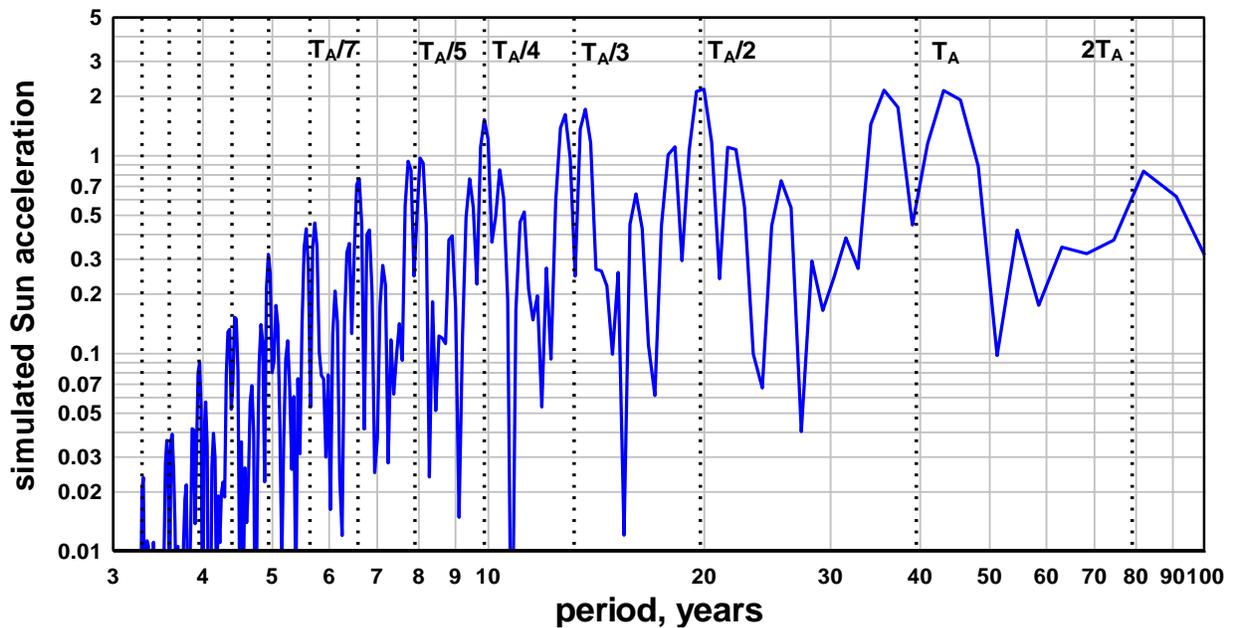

**Figure 8.** Spectral content of the simulated Sun – barycentre acceleration of Figure 7. A feature is alternating maxima and minor minima at $T_A/n$ periods. For example, a peak occurs at $T_A/2$ whereas a minor minima occurs between two peaks at $T_A/3$. A second feature is that secondary peaks occur between the dominant peaks. The secondary peaks occur at $T_A/(n+1/2)$, e.g. the peak at $T_A/4.5 = 8.8$ years

It is clear that the dominant peaks in the spectrum are associated with $T_A/n$. The peaks where n is even exhibit a peak centered on $T_A/n$ and two sidebands. The peaks where n is an odd number are split into pairs of peaks, for example the pairs of peaks centered at $T_A/3$, $T_A/5$ and $T_A/7$. It is also clear that secondary peaks will occur at $T_A/(n+1/2)$. The overall result of this section is that it is probable that planetary motion is the origin of the period $T_A$ = 39.5 years and that Sun acceleration relative to the barycentre is the origin of the $T_A/n$ spectrum of periodicities. This confirms the first objective of this paper, i.e., to ascertain that the period, $T_A$ = 39.5 years originates in Sun motion relative to the barycentre.

In a much more detailed analysis of Sun motion than provided here Cionco and Abuin (2016) studied the planetary torque exerted by the large planets on the Sun and showed that the spectral content of the torque in the sub-decadal spectral range of their study, periodicities in the discharge of large rivers, had dominant peaks at 6.56, 7.86 and 9.92 years. We note that the periods of planetary torque obtained by Cionco and Abuin (2016) correspond closely to the periods $T_A/6$, $T_A/5$ and $T_A/4$ years, (6.58, 7.90, and 9.87 years), respectively indicating that Sun torque and Sun acceleration have similar periodic content. Cionco and Abuin (2016) calculated the vectorial components of torque that are relevant to the spin-orbit coupling of the planets to the Sun. It is evident that the sharp peaks of torque are separated by ~ 39.5 years, c.f. Figure 9 and Fig 11, Cionco and Abuin (2016). The times coincide closely with the times of extreme Sun- barycentre acceleration in Figure 6.

A tentative explanation of why periodicity close to the harmonics of $T_A/n$ occurs in atmospheres in the solar system is as follows: The Sun is subjected to sharp impulses in acceleration or torque when the Sun passes in close proximity to the barycentre. The strongest impulses are separated by $T_A$ = 39.5 years. Assuming the impulses in Sun acceleration increase solar activity and the solar activity influences planetary atmospheres then, if the atmospheres on the Sun and on the planets have natural internal modes of oscillation, the modes of oscillation that are close in period to $T_A/n$, and possibly $T_A/(n+1/2)$, will tend to respond more strongly than other modes and will therefore appear with enhanced amplitude in the spectra of the atmospheric oscillations.

### 4. $T_A/n$ periodicity in the atmosphere of the Sun.

The magnetic field of the Sun varies with an approximately 22 year magnetic cycle known as the Hale cycle. Twice every Hale cycle the magnetic field becomes toroidal and sunspots emerge of the surface in a cycle called the Schwabe cycle. The variation of the solar atmosphere, or solar activity, can be measured by counting sunspots. Sunspots have been recorded accurately from 1700. Figure 9 shows the Group Sunspot Number (GSN), Svalgaard and Schatten (2016), and Figure 10 shows the periodogram obtained by FFT padded to 2048 years so as to better specify the periods of the long period peaks.

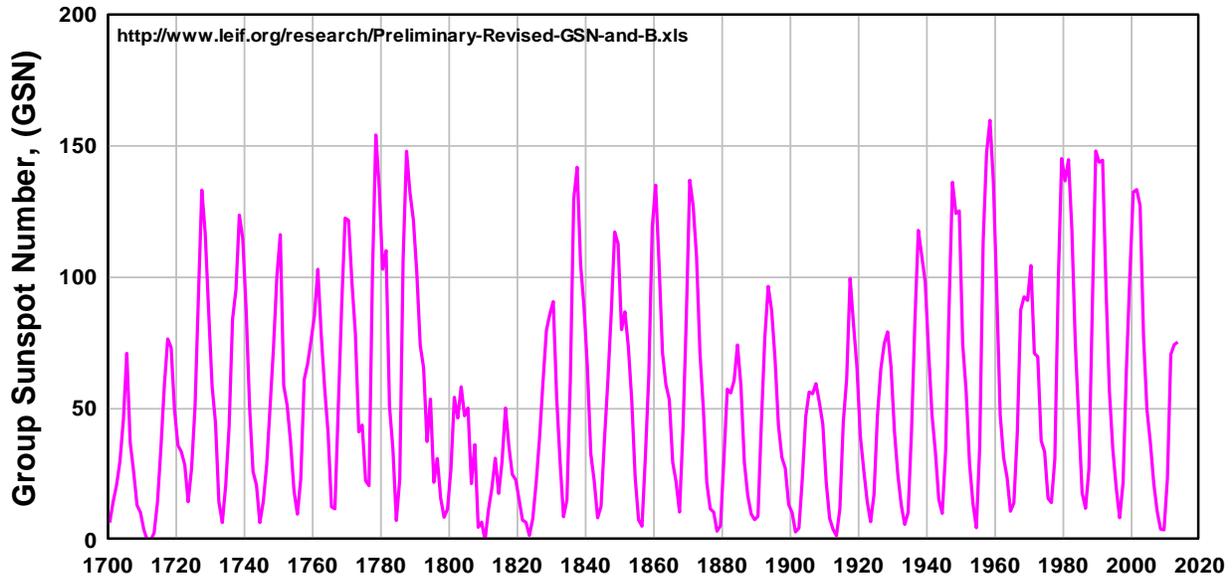

**Figure 9. The Group Sunspot Number (GSN) 1700 to 2013,** http://www.leif.org/research/Preliminary-Revised-GSN-and-B.xls

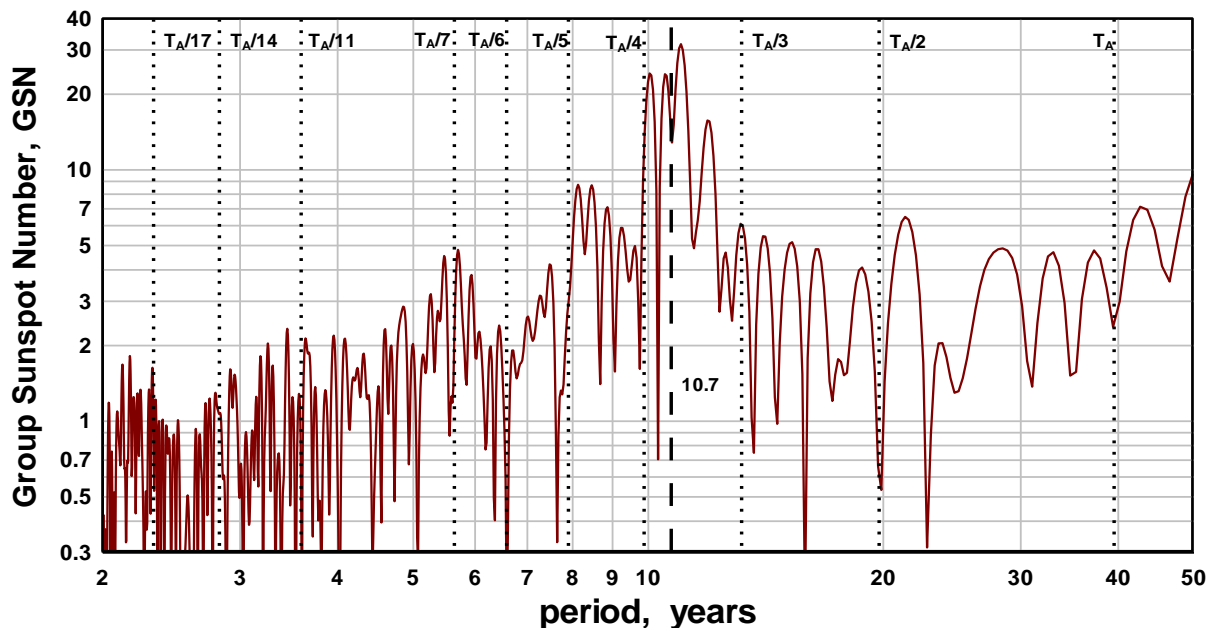

**Figure 10. The periodogram of the GSN. Some harmonics of $T_A$ = 39.5 years are indicated by reference lines and labeled with the corresponding $T_A/n$. Complex enhancements of amplitude are most noticeable at $T_A/5$, 7.90 years, at $T_A/7$, 5.64 years and at $T_A/4$, 9.87 years. The deep minimum at $T_A/6$ is also a noticeable feature. The enhancement around $T_A/17$, 2.32 years, corresponds in period to the solar QBO, Deng et al (2019). The average period of the Schwabe solar cycle, 10.7 years, is indicated by a dashed line.**

Figure 10 shows the nominal average period of the Schwabe cycle as the dashed line at 10.7 years. Several harmonics of $T_A/n$ are indicated with reference lines. While the spectrum is complex due to overlap with the Schwabe cycle at 10.7 years there is considerable variance of GSN evident at $T_A/3$, $T_A/4$, $T_A/5$, $T_A/7$ and $T_A/17$ that is clearly unrelated to the Schwabe cycle and corresponds reasonably closely in period with the periodicities of Sun motion as indicated in Figure 6 in the previous section. Figure 11 presents the region of the GSN periodogram between $T_A/3$ and $T_A/7$ with a scaled version of the periodogram of $R_B/R_{SUN}$ overlaid to facilitate comparison.

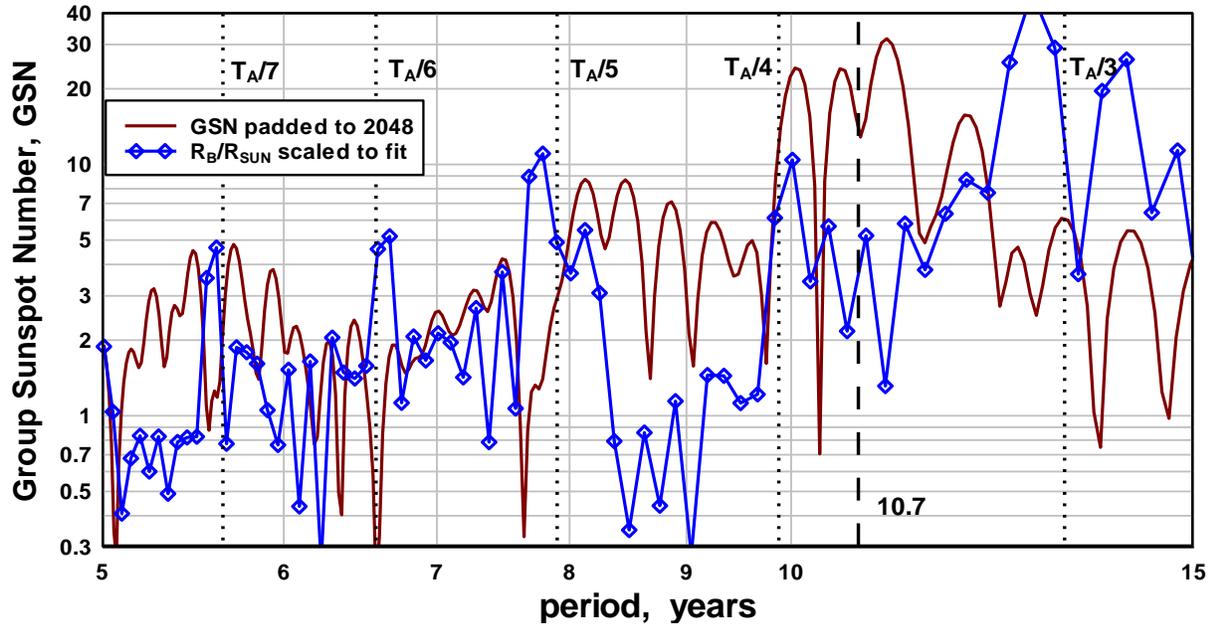

**Figure 11.** Compares the spectral content of GSN and $R_B/R_{SUN}$ (scaled for comparison) between 5 and 15 year period. A complex relationship is evident at $T_A/4$, $T_A/5$, $T_A/6$ and $T_A/7$.

The comparison in Figure 11 shows that the relationship between Sun motion and sunspot emergence is complex. However, as major features in $R_B/R_{SUN}$ and GSN occur at the same $T_A/n$ harmonic periods it is clear that a relationship does exist. The peak at $T_A/4$ in both $R_B/R_{SUN}$ and GSN appears to be unrelated to the average period of the Schwabe cycle, the 10.7 year reference line in Figure 11. The $T_A/4$, 9.87 year, periodicity is prominent in the PDO, c.f. Figure 1, and an interaction between this approximately 10 year periodicity in the PDO and the approximately 11 year Schwabe cycle has been discussed by van Loon and Meehl (2014). The periodicity at $T_A/5$, 7.9 years, coincides with the strongest periodicity in the NAO, c.f. Figure 1. A possible relationship between the ~7.9 year periodicity in GSN and the ~7.9 year periodicity in the NAO was discussed by Palus and Novotna (2009).

An interesting aspect of the comparison in Figure 11 is that the possible complex response of GSN to Sun motion evident at $T_A/7$, $T_A/5$ and $T_A/4$ leads to broad elevations of the amplitude of GSN at those harmonics of Sun motion. However, at the $T_A/6$ harmonic, period 6.58 years, the GSN response is depressed and does not show the broad elevation of amplitude that is evident at $T_A/7$, $T_A/5$ and $T_A/4$, respectively 5.64, 7.90 and 9.88 years. We show below, in Table 1, that the lack of response of GSN to Sun motion at $T_A/6$ is reflected in the absence of variance at period $T_A/6$ in other atmospheres in the solar system. We can immediately appraise this lack of response at $T_A/6$ in the comparison of the NAO and PDO spectra in Figure 1. Whereas variance in the NAO or the PDO is prominent at $T_A/4$, $T_A/5$ and $T_A/7$, neither the NAO nor the PDO show significant variance at $T_A/6$. This leads to the conclusion that, most likely, any cause and effect sequence is from Sun acceleration to solar activity and from solar activity to planet surface oscillations. Thus, for example, the period of the QBO on Earth at ~ 28 months would derive from the period of the QBO on the Sun at ~ 28 months, (Zaqarashvili et al 2010, Deng et al 2019, Inceoglu et al 2021).

There has been considerable research on how the motion of the planets might influence solar activity. For example, Abreu et al (2012) compared the periodicity of planet induced torque on the solar tacholine with the periodicity of reconstructed solar activity over the past 9000 years and showed there were coincidences in the 100 year to 500 year period range. The Abreu et al (2021) study has received support, Charbonneau (2010), and criticism, (Cameron and Schussler 2013, Polianov and Usoskin 2014). However, the 100 to 500 year period range is well outside the range of interest here, a range that is limited by the length of the observational record of planetary atmospheres to below the 100 year period range. It may be relevant that the slow magnetic Rossby waves in the solar tacholine that modulate solar cycle strength exhibit a spectrum of periodicity, given by $T_R = -4700/(2 – p(p+1))$, years, Zaqarashvili et al (2015). Here p is the poloidal wave number of the Rossby waves. For example, when p = 24, $T_R$ = 7.9 years, ~ $T_A/5$ years. This spectrum of solar internal periodicity overlaps the spectral range of interest in this paper.

## 5. $T_A/n$ periodicity in planet atmospheres

The relationship between solar activity and large scale climatic phenomena like the NAO, the PDO and the SOI has been studied by Velasco and Mendoza (2008) who established the periodicity range in which significant coherence between large scale climate phenomena and solar activity occurred. However, this study did not consider the relationship of climate phenomena to Sun motion. The relationships of surface temperature to Sun motion, Charvatova and Strestik (2004), and rainfall to Sun motion, Cionco and Abuin (2016), have been studied previously. However, the relationship of large scale climatic phenomena like the NAO, PDO and SOI to Sun motion, has, to the author's knowledge, not been studied. It is interesting to note that Charvatova and Strestik (2004) found a similarity in the spectra of surface air temperature and solar motion at periods, 12.8, 10.4 and 7.8 years, close respectively to $T_A/3$, 13.1 years, $T_A/4$, 9.9 years, and $T_A/5$, 7.9 years. And Cionco and Abuin (2016) found significant spectral lines in river discharge at 6.5, 7.6, 8.7 and 10.4 years, close respectively to $T_A/6$, 6.6 years, $T_A/5$, 7.9 years, $T_A/4.5$, 8.8 years and $T_A/4$, 9.8 years.

**5.11 $T_A/n$ periodicity in the SOI**   The Southern Oscillation is the most dominant mode in short term climate variation over the globe, accounting for a significant portion of the variance in the global climate system. The state of the Southern Oscillation is given by the Southern Oscillation Index (SOI), a measure of the sea level pressure difference between Tahiti and Darwin, (Chu and Katz 1989, Trenberth and Jones 2007). The monthly and annual average SOI, 1876 to 2023, is shown in Figure 12. Also shown is the $T_A/11$, 3.59 year period component of the SOI which is discussed below. The $T_A/11$ component is the strongest mid frequency component in the spectrum of the SOI, see Figure 13.

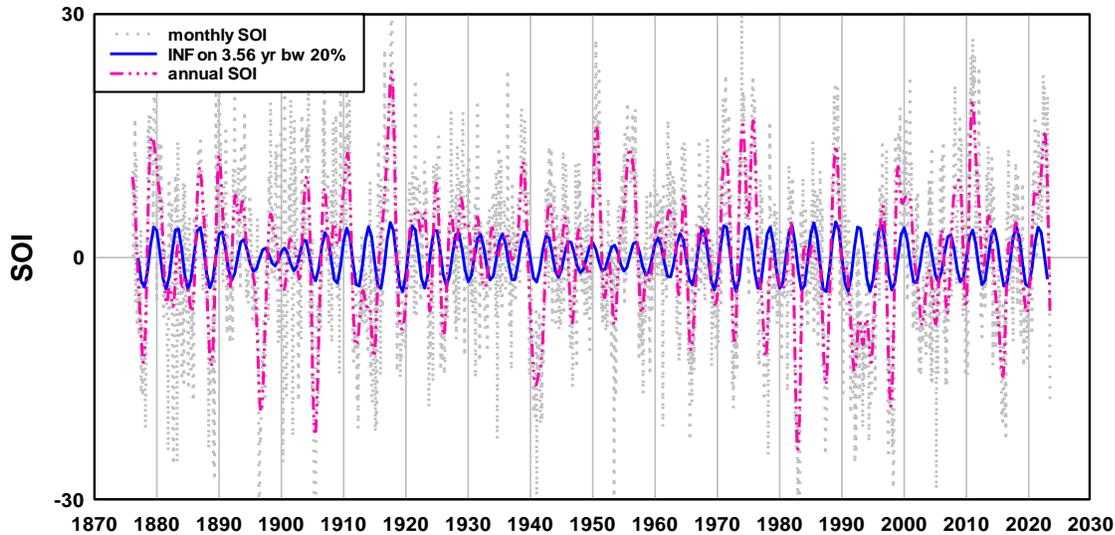

**Figure 12. The monthly and annual SOI variation, 1876 to 2023. Also shown, the narrow band filtered (INF) component centered on 3.59 years, $T_A/11$, for 20% bandwidth of the inverse notch filter.**

Figure 13 shows the FFTs of the first and second half of the SOI record and the average FFT. This shows that persistent periodicities, i.e. periodicities that persist over most of the record in the SOI, are centered close to the $T_A/7$, $T_A/11$, $T_A/14$ and $T_A/17$ periods, respectively the 5.64, 3.59, 2.82 and 2.32 year periods. The sharp peak at $T_A/17$, 2.32 years, is associated with the QBO. However the other persistent peaks are rather broad. Obtaining the FFTs for the first, second, third and fourth quarter of the SOI, Figure 14, provides further insight on why the peaks are broad.

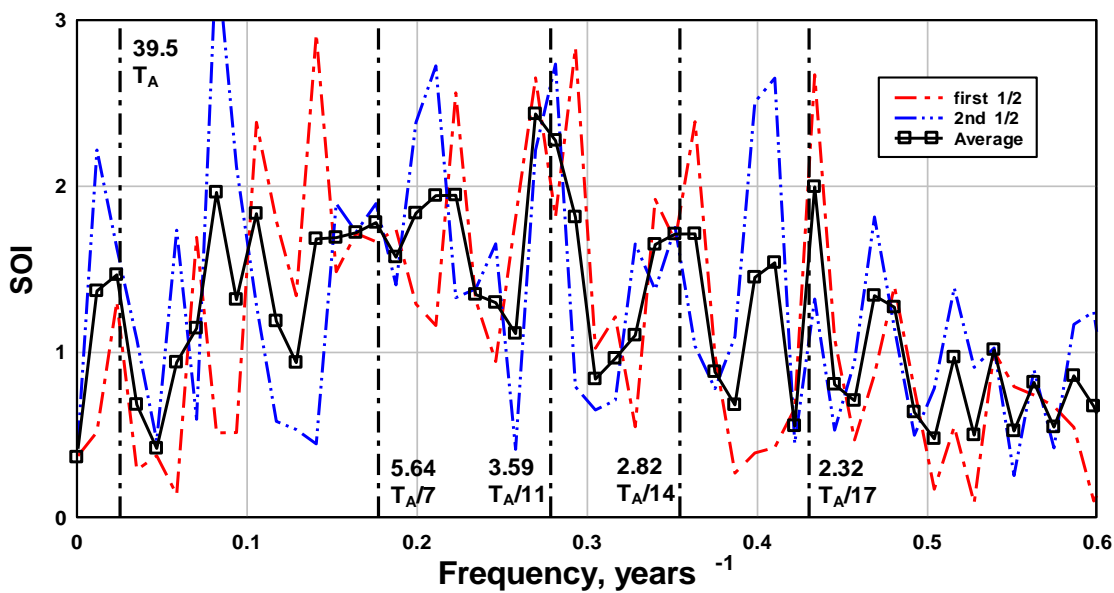

**Figure 13. The spectra from the first and second halves of the SOI record. The average spectrum is also shown. The dominant peak occurs at $T_A/11$, 3.59 years. The peak at $T_A/17$, 2.32 years, is due to teleconnection with the QBO.**

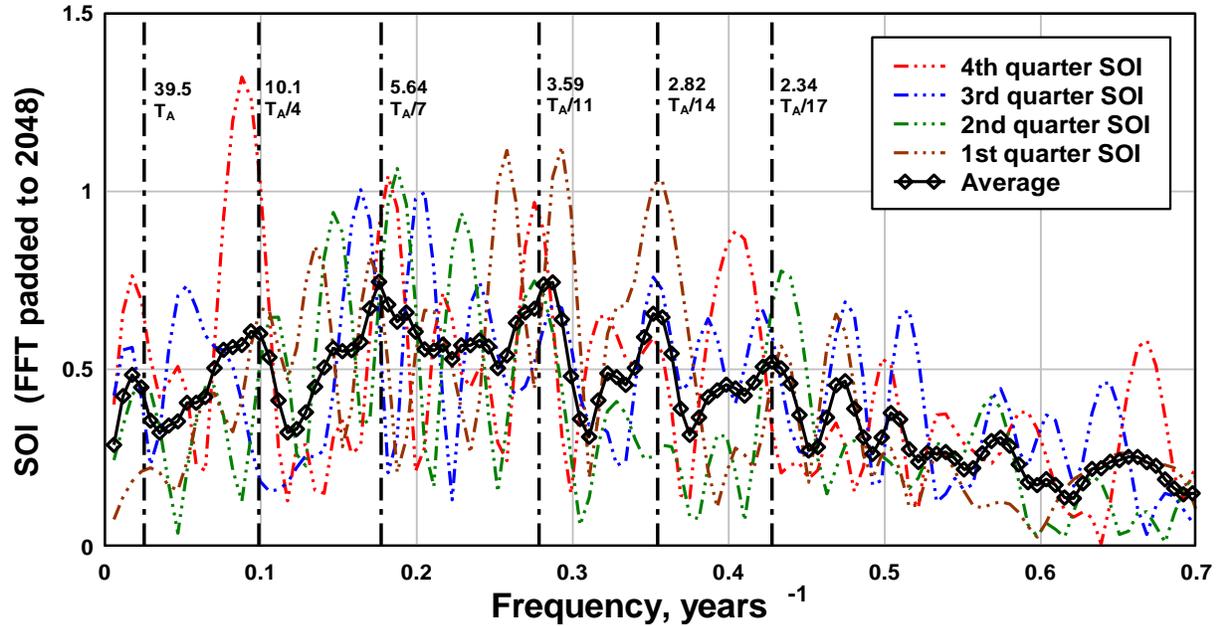

**Figure 14.** The spectra from the first, second, third and fourth quarters of the SOI record. The average spectrum is also shown. The periodicities vary from quarter to quarter but cluster around the $T_A/n$ periods marked by reference lines as is evident in the average spectrum.

In Figure 14, in order to a better estimate of the periods of the resolved peaks, each quarter segment 440 months long, was padded with zeroes to 2048 months. It is obvious from Figure 14 that the periodicity from one quarter to the next is variable. However, the average of the four spectra shows the periodicities are clustered around the $T_A/n$ harmonics indicated by reference lines. The periodogram of GSN, Figure 11, shows that the variance at $T_A/7$ in GSN is associated with four closely spaced peaks. It is tempting to speculate that the cluster of peaks around the $T_A/7$ period in the SOI may be due to the oscillations of the SOI, responding to periodic solar forcing associated with the cluster of periodicities close to $T_A/7$ in the GSN spectrum.

**5.12 Time variation of the $T_A/11$ component of the SOI.** While an objective of this paper is to show that dominant oscillations of planet atmospheres have periods close to $T_A/n$, and this objective can be achieved by spectral analysis alone, it is useful to look briefly at the time dependence of atmospheric variation. The $T_A/11$, 3.59 year, component is the dominant mid frequency component of SOI variation, c.f. Figure 13. Comparisons of the time variation of different types of variables is difficult as time variations depend on phase, and as there is usually some delay between cause and effect, some estimate of the delay must be made to facilitate any comparison. Here we compare three time variations: The $T_A/11$ components of Sun motion using $R_B/R_{SUN}$, solar activity using the GSN, and atmospheric oscillation using the SOI. Figure 15 shows the $T_A/11$, 3.59 year, component of $R_B/R_{SUN}$. Figure 16 compares the variation of this component with the variation of the $T_A/11$ component of GSN. To facilitate the comparison both variations have been normalized by division by standard deviation. To further facilitate the comparison the GSN component has been advanced by one year to bring the two variations into phase. That is, a delay of one year between effect of Sun motion and the resultant solar activity maximizes the correlation between the two components.

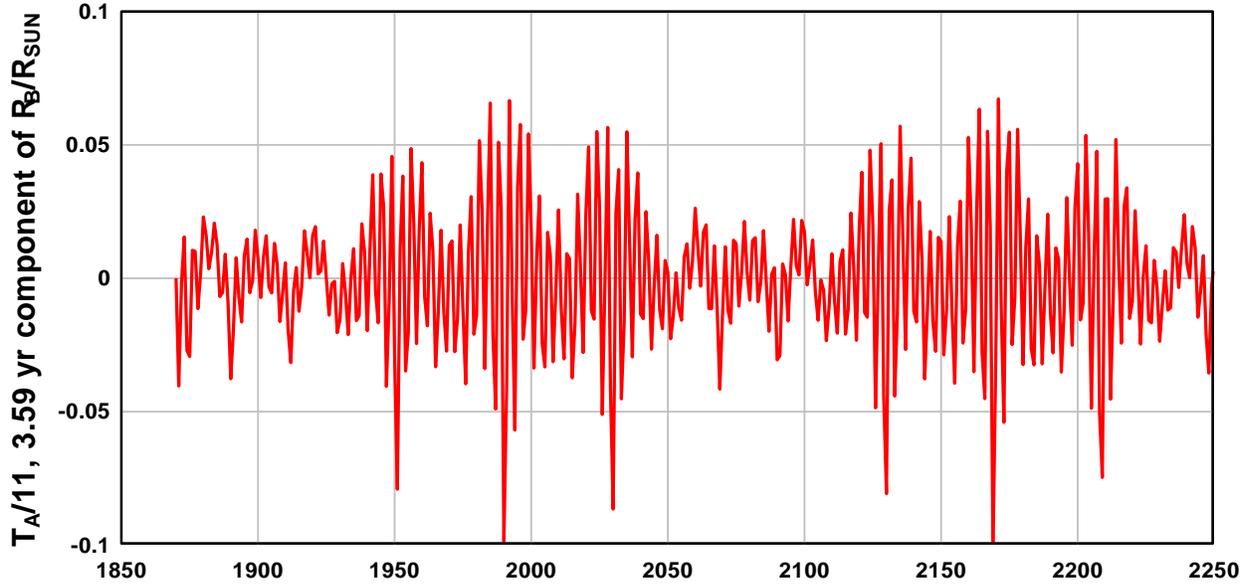

Figure 15. The $T_A/11$ component of $R_B/R_{SUN}$ obtained by INF at 5% band width.

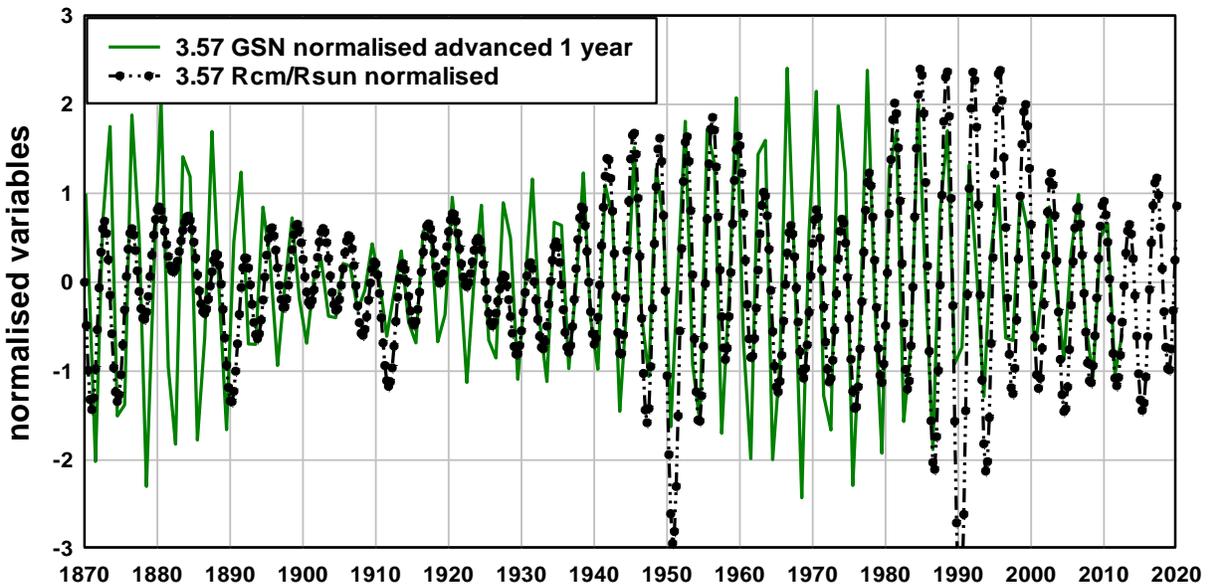

Figure 16. Compares normalized values of the $T_A/11$ components of $R_B/R_{SUN}$ and GSN. The GSN component has been advanced by one year. The correspondence obtained suggests that the variation in the $T_A/11$ component of GSN may be delayed by one year with respect to the $T_A/11$ component of $R_B/R_{SUN}$.

With a one year delay between solar motion and solar activity included the phase relation is seen to be stable in phase over the entire time interval 1870 to 2020, the interval of the SOI record. Figure 17 compares with $T_A/11$ components of the GSN and the SOI without adjustment to the phase of the components, i.e. it is assumed that the cause and effect delay between solar activity and SOI variation is insignificant in respect to the ~ 3.6 year period variations being compared. It is evident from Figure 17 that the $T_A/11$ component of SOI suffers a 180 degree phase shift relative to the GSN component at around 1900.

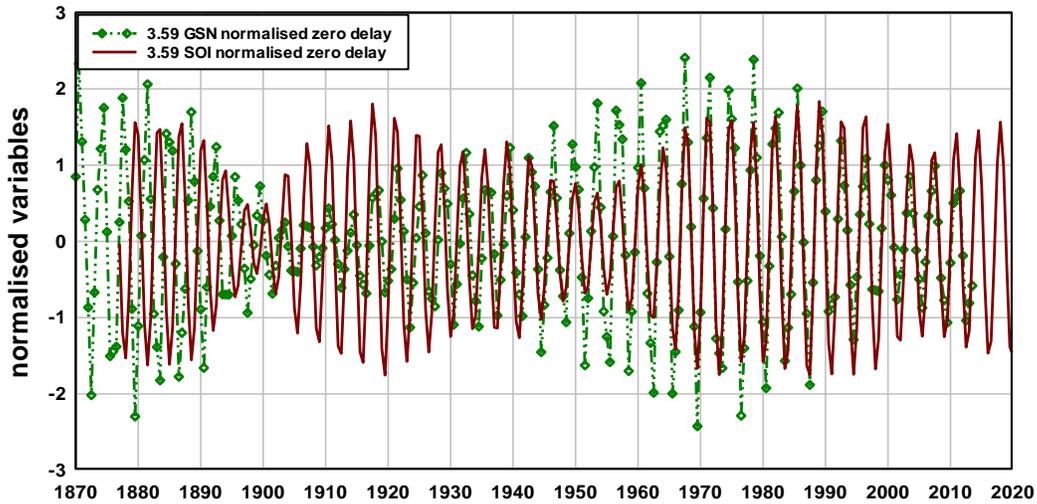

**Figure 17. Compares the $T_A/11$ normalised components of GSN and the SOI. A phase shift of the SOI component relative to the GSN component is evident at around 1900. As the GSN component appears to be stable in phase relative to the $R_B/R_{SUN}$ component in Figure 16 it is assumed the phase shift evident here is due to the SOI component shifting in phase or varying in frequency at around 1900.**

From 1876 to 1900 the $T_A/11$ component of the SOI is in anti-phase to the $T_A/11$ component in GSN. After 1900 the two components are consistently in phase. We are not here interested in interpreting the phase relationship between solar activity and the SOI as a whole. However, it is interesting to assess the phase shift between the two narrow band components at 1900 in terms of a possible shift from one $T_A/n$ component of the SOI to another as this is an effect that appears to have occurred in the atmosphere of Jupiter and will be discussed in section 5.5. The apparent phase shift of the SOI component at 1900 is examined in more detail in Figure 18.

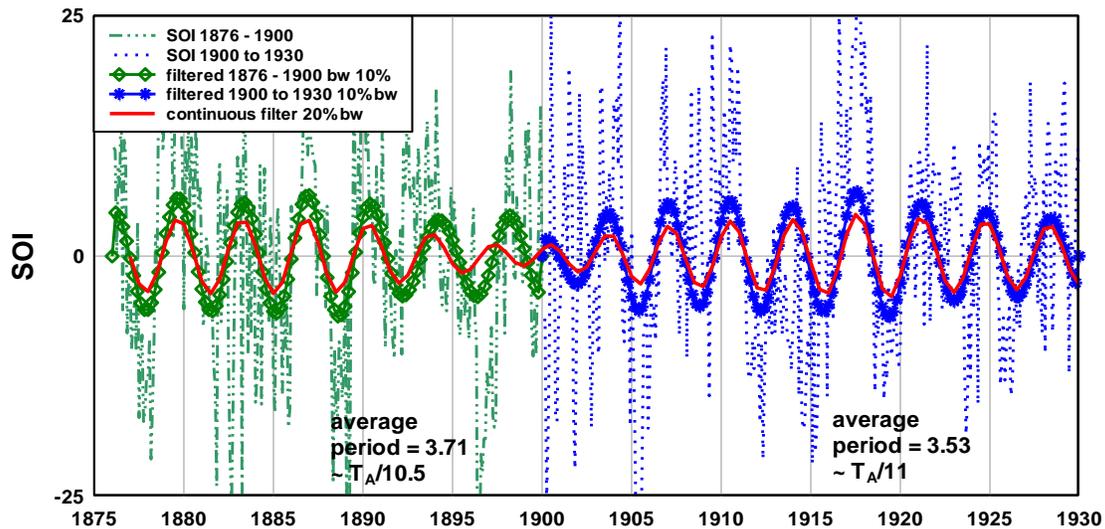

**Figure 18. The segment of the SOI record before 1900 and the segment for 30 years after 1900 are shown and analyzed separately. The average periods of each segment were obtained by counting peaks in each segment. The full red line is obtained by narrowband INF over the combined segments.**

Figure 18 shows the SOI record between 1876 and 1930. The first part of the record, 1876 to 1900, was filtered for the $T_A/11$ component by an INF filter with 10% band width, shown as green. The second part, 1900 to 1930, was similarly filtered, shown as blue. The average period in the 1876 to 1900 segment, obtained by cycle counting is 3.71 years, ~ $T_A/10.5$ years. The average period in the 1900 to 1930 segment is 3.53 years, ~ $T_A/11$ years. Thus it appears this component of the SOI transitioned in period by one half a $T_A/n$ state around 1900. The full red line in Figure 18 is obtained with an INF filter of centre period $T_A/11$ and 20% band width and replicates the transition of the SOI shown in Figure 17.

**5.2 $T_A/n$ periodicity in the NAO.** The spectra of the NAO and the PDO have been compared in Figure 1. Here we provide further detail of the NAO focusing on the periodicities that are persistent. Figure 19 shows the spectrum obtained from the entire NAO record with reference lines at $T_A/n$ to illustrate the correspondence of spectrum peaks and $T_A/n$ harmonics where correspondence occurs. Figure 20, showing FFTs for the first and second half of the NAO record, indicates that a number of periodicities are persistent over the entire record, e.g. periodicities at $T_A/5$, 7.9 years, and $T_A/8$, 4.9 years.

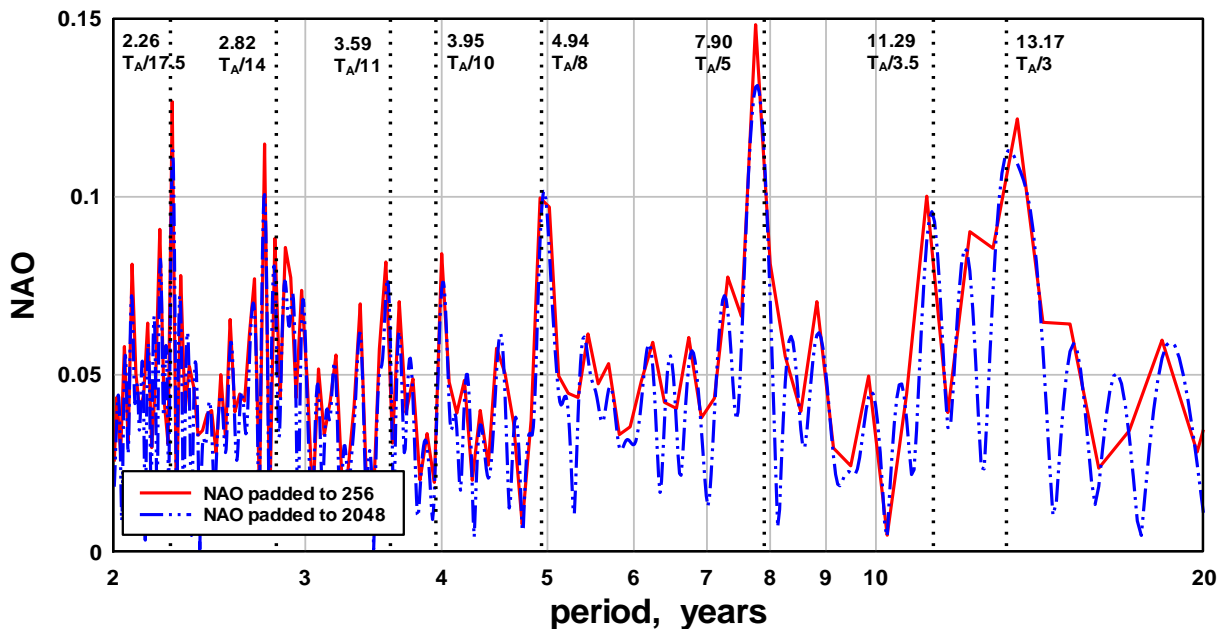

**Figure 19. The spectral content of the NAO in the period range 2 to 20 years. The $T_A/n$ periods close to the most prominent peaks are indicated with reference lines.**

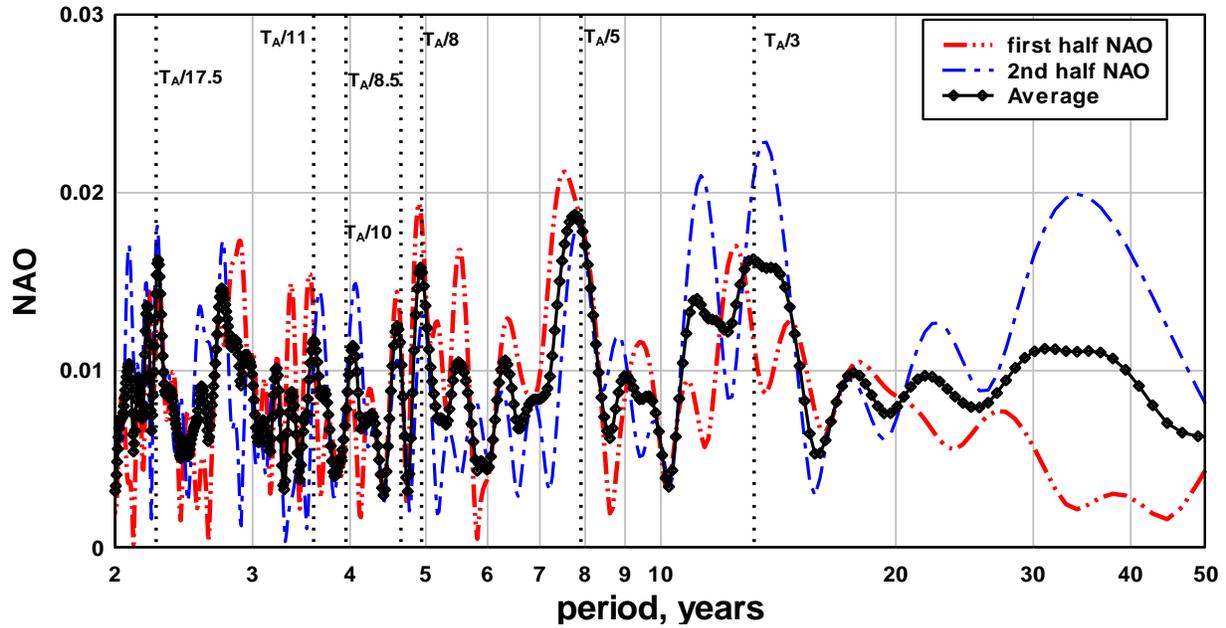

**Figure 20.** Compares the spectral contemt of the first and second halves of the NAO record and the average spectrum.

**5.3 $T_A/n$ periodicity in the IOD and the AO and AAO** The Indian Ocean Dipole, (IOD), is an oscillation of sea surface temperature between the western and eastern Indian Ocean. Positive/negative IOD results in drought/flood in S. E. Asia and Australia and flood/drought in East Africa, (Saji et al 1999, Webster et al 1999). There are about four each positive – negative events every 30 years, i.e. the IOD has an average periodicity of about 7.5 years. The spectral content of the first and second halves of the IOD record are compared in Figure 21 along with the average spectrum. Consistently enhanced spectral content is evident at five $T_A/n$ harmonics and at one $T_A/(n+1/2)$ harmonic, $T_A/1.5 = 26.3$ years, as marked by reference lines in Figure 21.

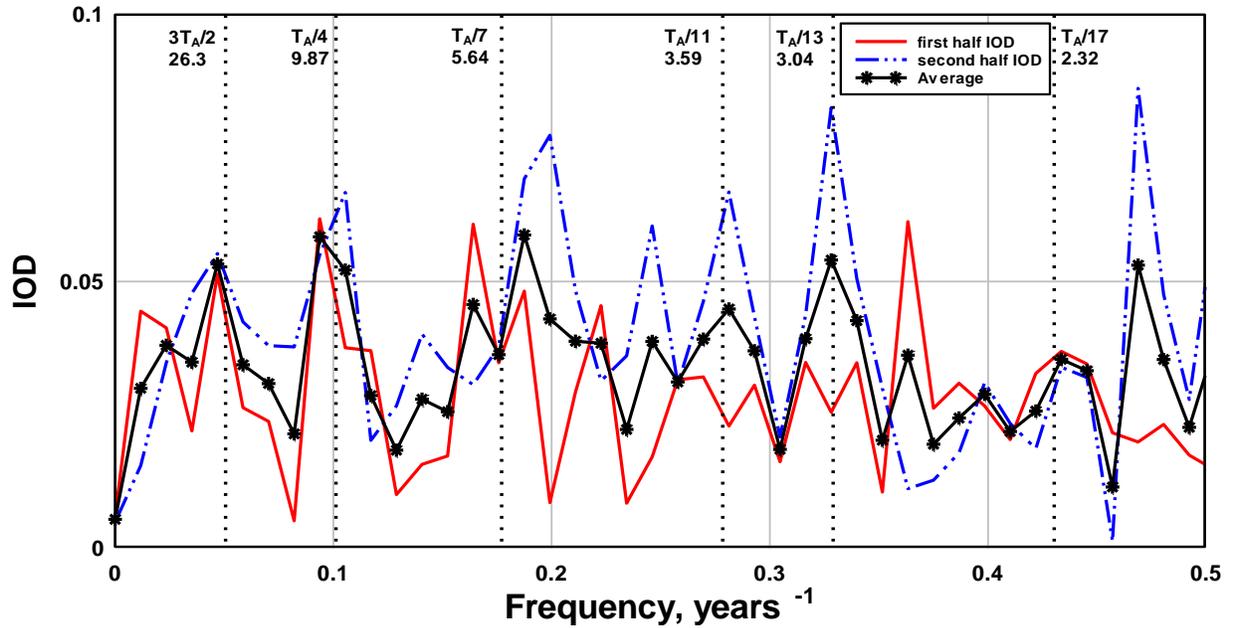

**Figure 21. The FFT spectra for the first and second half of the IOD, 1870 to 2023, and the average spectrum. The two consistently strong periods are close to $T_A/4$, 9.87 years, and $T_A/7$, 5.64 years.**

The Arctic and Antarctic Oscillations, AO and AAO, also known as the Northern and Southern Annular Modes, NAM and SAM respectively, are belts of high and low pressure anomalies at high latitudes. The belts oscillate at mid frequency between a negative state of extreme anomalies to a positive state of less extreme anomalies, Limpasuvan and Hartmann (1999). The record of the AO is 70 years long and the AAO record is 44 years long so it is difficult to accurately resolve periodicities longer than 10 years. Assuming the annular modes respond to a common forcing the spectra of both the AO and AAO were averaged to identify persistent periodicities, Figure 22.

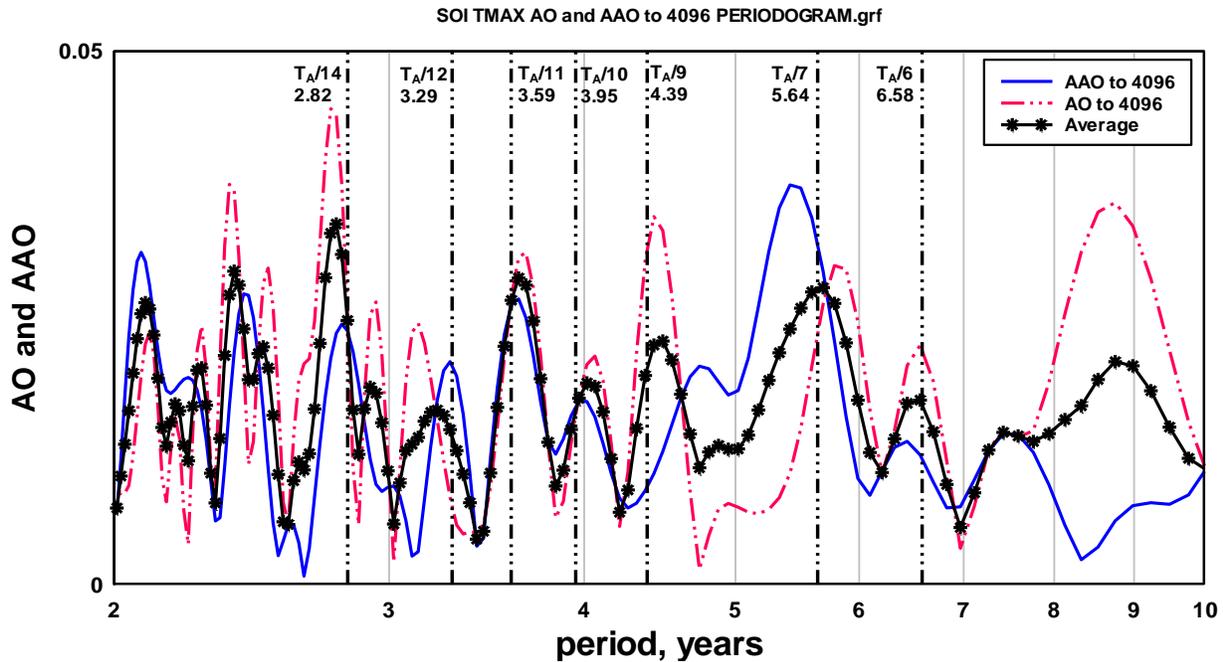

Figure 22. Shows the periodicity in the AO and in the AAO and the average periodicity. Although the AO and AAO records are derived from different areas of the globe, the north and south annular modes are expected to have similar natural periods and it is expected that the spectral response to external forcing would be similar and, therefore, averaging the spectra would be appropriate and useful.

The periods of the two strongest peaks in the average spectrum, Figure 22, occur close to $T_A/11$ and $T_A/14$ years. When the components at $T_A/11$, 3.59 years and $T_A/14$, 2.82 years are obtained by INF it is apparent that the $T_A/11$ components of AO and AAO oscillate in-phase and the $T_A/14$ components of AO and AAO oscillate in anti-phase, Figures 23 and 24.

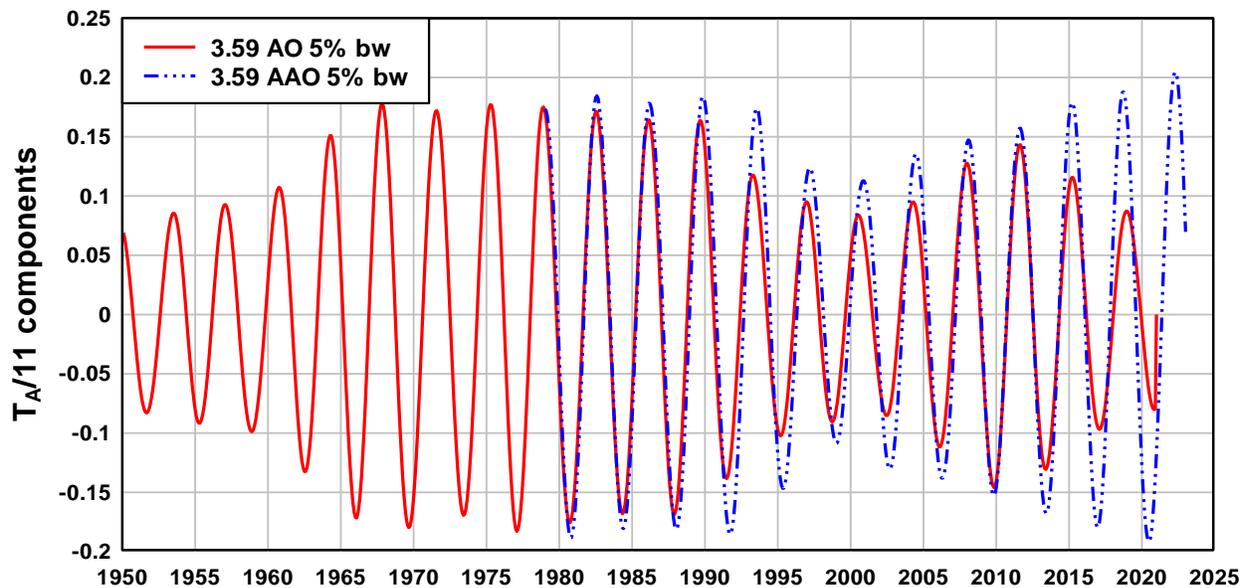

Figure 23. The $T_A/11$, 3.59 year period, components of the AO and the AAO.

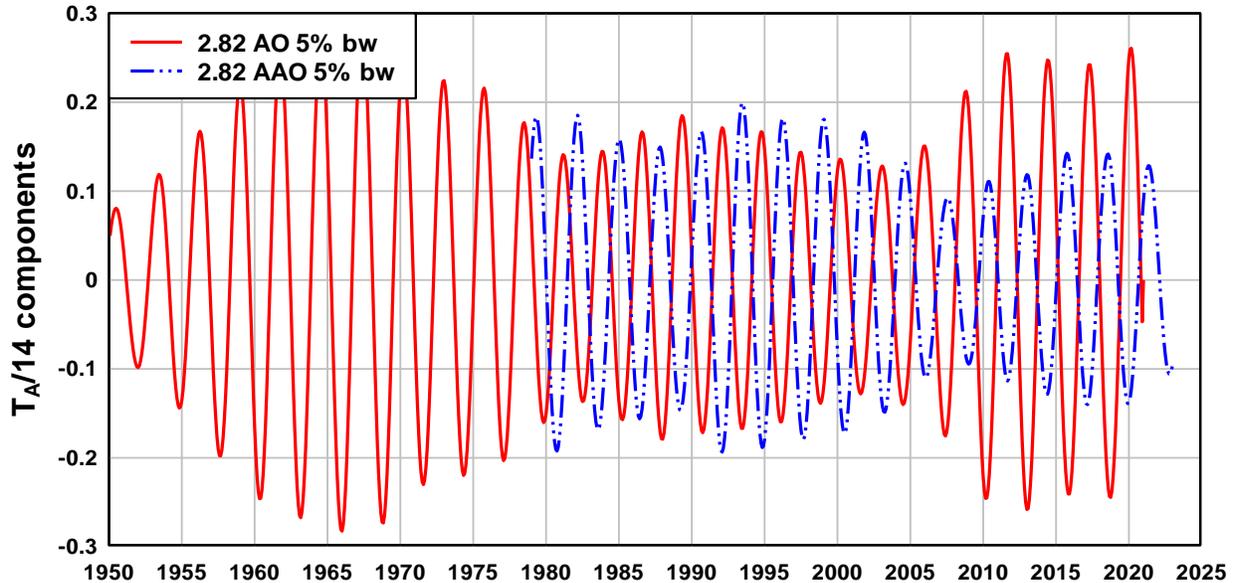

**Figure 24.** The $T_A/14$, 2.82 year period, components of the AO and the AAO.

**5.4 $T_A/n$ periodicity in the Quasi-Biennial Oscillation (QBO)** In the tropical stratosphere, deep layers of eastward and westward winds encircle the globe. The speed of each wind is about 20 m/s and the direction of the winds reverse two times, on average, every 28 months, or 2.33 years. The QBO is the most predictable mode of atmospheric variability not linked to the annual seasons. However, "the QBO is unrelated to any known periodic forcing and is therefore difficult to understand", Dunkerton et al (2015). The QBO affects climate phenomena like the NAO and the Madden-Julian Oscillation, Anstey et al (2022). A theory of the QBO was developed by Lindzen and Holton (1968) who showed that the QBO was driven by a broad spectrum of vertically propagating, short period (5 – 15 day period), waves excited in the troposphere. Theory and observations indicate that a broad spectrum of vertically propagating waves must be considered to explain the QBO, (IPCC 2007, Anstey et al 2022). The QBO anomaly, 1948 – 2023, is shown in Figure 25 along with the $T_A/17$, (2.32 year), component obtained by INF. Figure 26 shows the FTT of the QBO and the FFT of the $T_A/17$ year component.

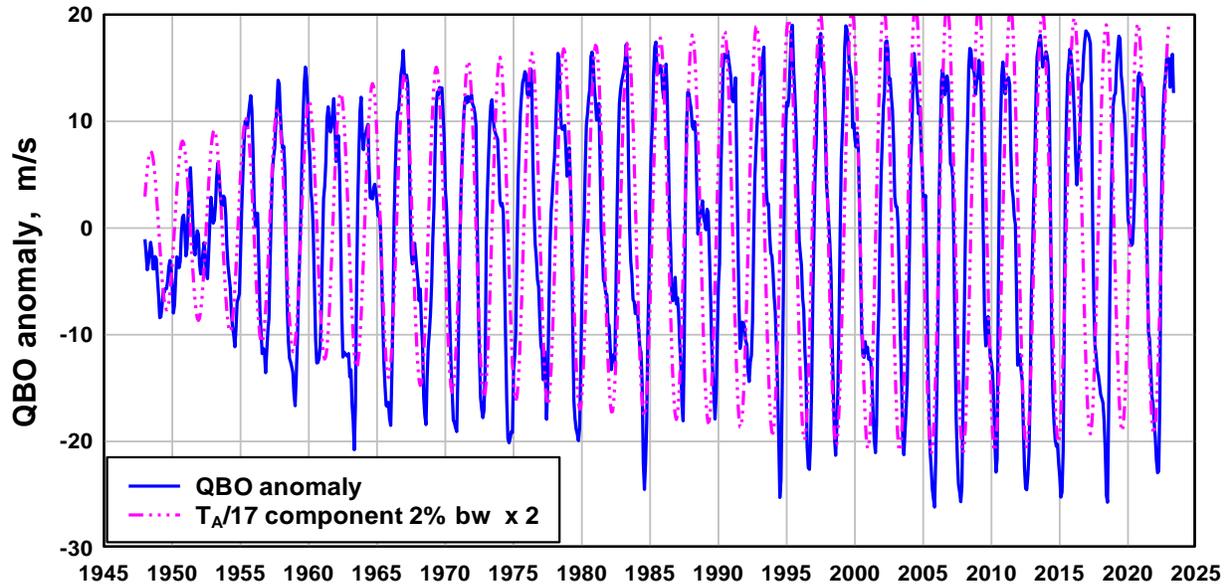

**Figure 25. The monthly QBO anomaly, 1948 – 2023 and the $T_A/17 = 2.32$ year component determined by INF. Anomalies in the sinusoidal pattern of variation are evident around 1950, 1965, 1990, and 2015.**

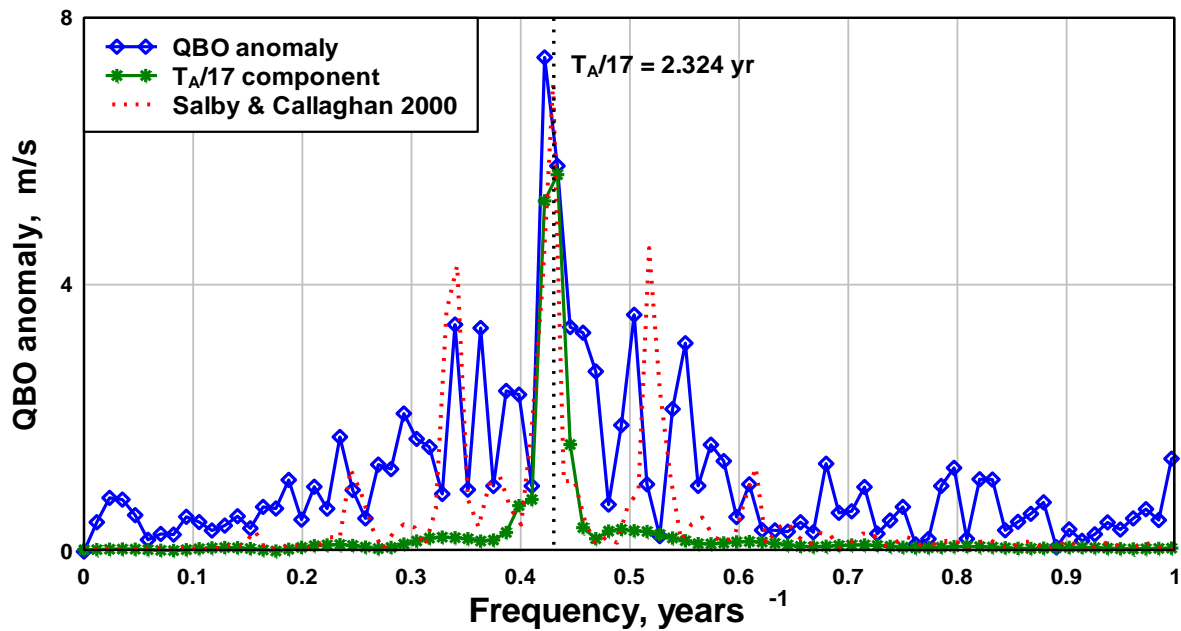

**Figure 26. Although the QBO variation, Figure 25, appears to be almost monochromatic the spectral content has a large contribution at sidebands to the main period. The spectrum of the QBO model of Salby and Callaghan (2000) which is based on frequency modulation of the QBO by the 11 year solar cycle is shown for comparison.**

It is evident from the spectrum in Figure 26 that there is considerable variability in the QBO at periods significantly different from $T_A/17$, i.e. the QBO is not a simple harmonic oscillator driven by a source of single period $T_A/17$. Various workers have attempted to model the QBO variation and obtain a fit to the spectrum. For example, Salby and Callaghan (2000) modeled the QBO as a simple harmonic oscillator frequency modulated by the 11-year solar cycle. The red dotted curve in Figure 26 shows the spectrum of an oscillator of period 2.32 years that is frequency modulated by an 11 year period cycle as proposed by Salby and Callaghan (2000). The frequency modulated model spectrum is symmetrical in frequency about the centre frequency with sideband peaks at frequencies spaced at n/11 years$^{-1}$. However, when the QBO spectrum is obtained with the record padded to 4096 months, Figure 27, it is clear that the QBO spectrum is more consistent with a more finely spaced sideband structure than that due to an 11 year modulation. The reference lines in Figure 27 are spaced at n/$T_A$ years$^{-1}$, about one quarter the spacing due to an 11 year modulation.

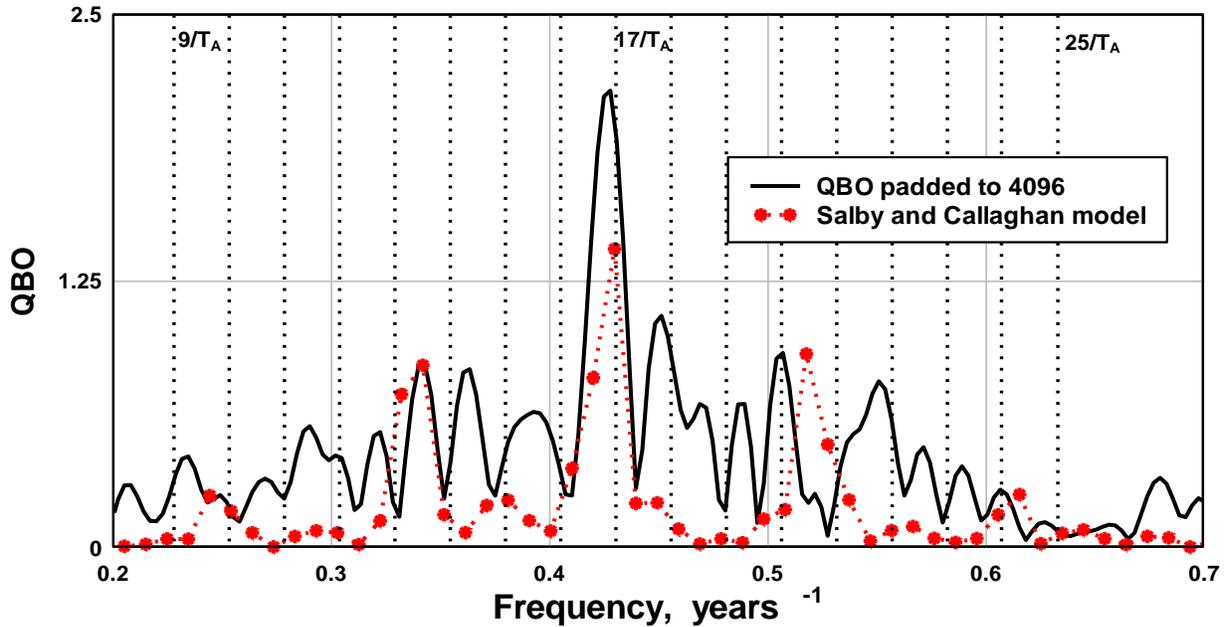

**Figure 27. Provides better definition of the resolved peaks in the QBO spectrum by obtaining the spectrum after padding the record to 4096 months and then obtaining the FFT. Sideband peaks with spacing more consistent with spacing 1/$T_A$ yr$^{-1}$ then become apparent. The spectrum associated with the Salby and Callaghan (2000) model, where sideband peak spacing is 1/11 yr$^{-1}$, is also shown for comparison.**

There is considerable regularity in the spectral content of the QBO outside the main peak at 17/$T_A$ years$^{-1}$. The stronger peaks at frequencies above 17/$T_A$ occur close to 18/$T_A$, 20/$T_A$ and 22/$T_A$. For peaks below the frequency of the main peak at 17/$T_A$ the peaks all occur at frequencies where n takes half integer values, e.g. the peak at 13.5/$T_A$. Intuitively, one would expect a spectrum as in Figure 26 or Figure 27 if the QBO was a simple harmonic oscillator forced by a broad spectrum source with a range of periodicities similar to that due to Sun acceleration, the periodogram of which is shown in Figure 6. In Section 6 we model just such a forcing and show that the spectrum of the modeled response is indeed similar to the QBO spectrum.

**5.5 $T_A$/n periodicity in the atmospheres of Jupiter, Saturn and Neptune.** The equatorial stratosphere of Jupiter exhibits a periodic oscillation of temperature and winds with height. Jupiter's Quasi-Quadrennial Oscillation (QQO) has recently been observed to experience a transition from one periodic state to another quite different periodic state, Antuano et al (2021).

Jupiter's stratospheric temperatures between 1980 and 2011 show two significantly different periods for the QQO with a 5.7-year period (c.f., $T_A/7$ = 5.64 years) between 1980 and 1990 and a 3.9-year period (c.f., $T_A/10$ = 3.95 years) between 1996 and 2006, Antuano et al (2021). Thus, the observations of Jupiter's atmosphere indicate that around 1990 a disruption resulted in the QQO transitioning from ~$T_A/7$ period to ~$T_A/10$ period. A disruption of the QQO around 1990 would closely coincide with the occurrence of a strong peak in Sun acceleration, c.f. Figure 5. Antuano et al (2021) mention a second disruption to the QQO pattern occurring around 2007, a time consistent with another peak in Sun acceleration, c.f. Figure 5.

The period of Saturn's equatorial stratospheric oscillation is as yet not well defined. Either the oscillation has a period close to ~20 years or its period is naturally variable, Blake et al (2023), and the oscillation is occasionally interrupted as in 1990 when the equatorial stratosphere showed strong thermal perturbations. If further observation confirms a period of ~20 years the period would correspond closely to $T_A/2$ (= 19.75 years). While also not well defined, a ~ 13 year period oscillation is observed in Neptune's cloud cover, Chavez et al (2023), a periodicity close to $T_A/3$ = 13.16 years.

A possible reason for disruptions of planet atmosphere oscillation coinciding with peaks in Sun acceleration is discussed in the next Section.

## 6. Model of a QBO type oscillator forced by periodic acceleration impulses.

Here we develop a model of a QBO type oscillator as follows: We assume the atmosphere in question, the atmosphere of the Sun or the atmosphere of a planet, behaves as a mass elastically suspended and experiencing some form of damping, i.e. behaves as a damped simple harmonic oscillator (SHO) similar to the functioning of a seismometer. In response to peaks in Sun acceleration the atmosphere will respond to that acceleration in a manner similar to the acceleration impulse response of a SHO. The sharply peaked time variation of Sun acceleration relative to the barycentre, is shown in Figure 5. The acceleration variation in Figure 5 is somewhat complex due to the different spacing and amplitude of the peaks. In the simplified model developed here the acceleration is assumed to have constant amplitude spikes with a constant spacing of 10 years. The spikes are further simplified to impulses in time (delta functions, $\delta(t)$), so that the time response is given by the Green's function for the impulsive response of a SHO:

$G(t) = H(t)\exp(-\beta t)\sin(\omega_i t)/\omega_i$

where $\omega_i = (\omega_0^2 - \beta^2)^{1/2}$, $\omega_0 = 2\pi/T_0$, $T_0$ is the natural period of the SHO and $\beta$ is the damping constant. H(t), the Heaviside step function, is 0 before the impulse and 1 after. For this model $T_0$ is taken as 2.32 years and $\beta$ = 0.05. As the damping is small $\omega_i$ can be approximated as $\omega_0$. The responses of the SHO to each impulse applied at intervals of $T_M$ = 10 years is shown in Figure 28. When the responses to each impulse are summed the overall response is as shown in Figure 29. The periodogram of the overall response is shown in Figure 30.

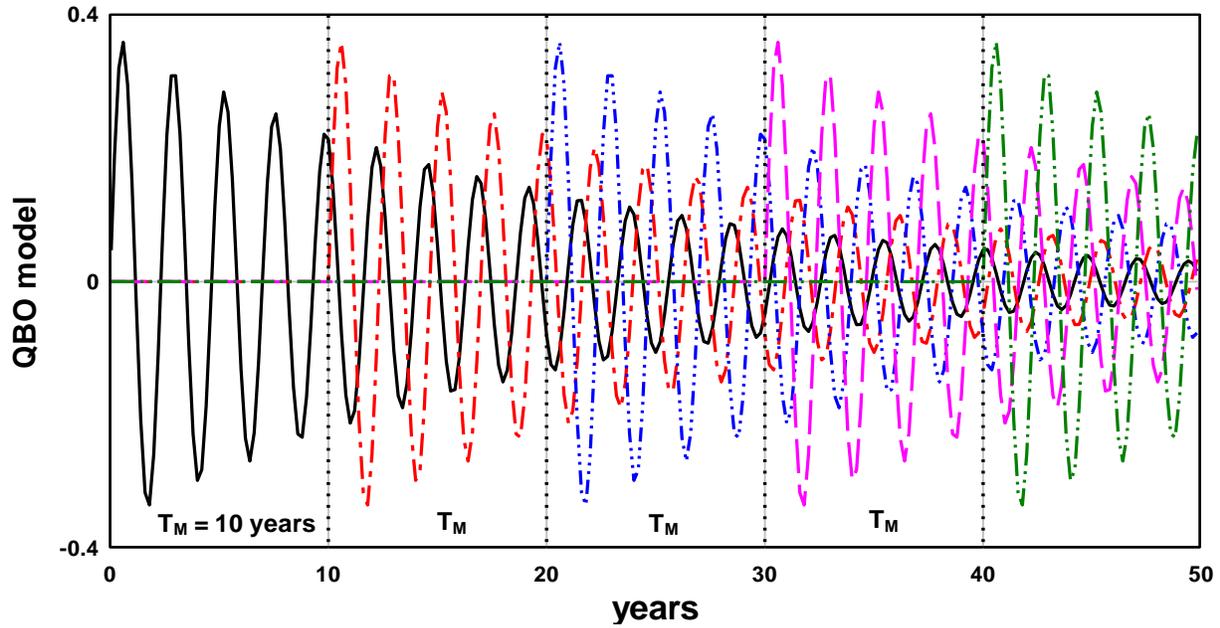

Figure 28. The impulse response of a damped simple harmonic oscillator of natural period $T_0 = 2.32$ years when acceleration impulses are applied at intervals of $T_M = 10$ years can be obtained by summing the sequence of responses, shown in this Figure, to each individual impulse, see Figure 29.

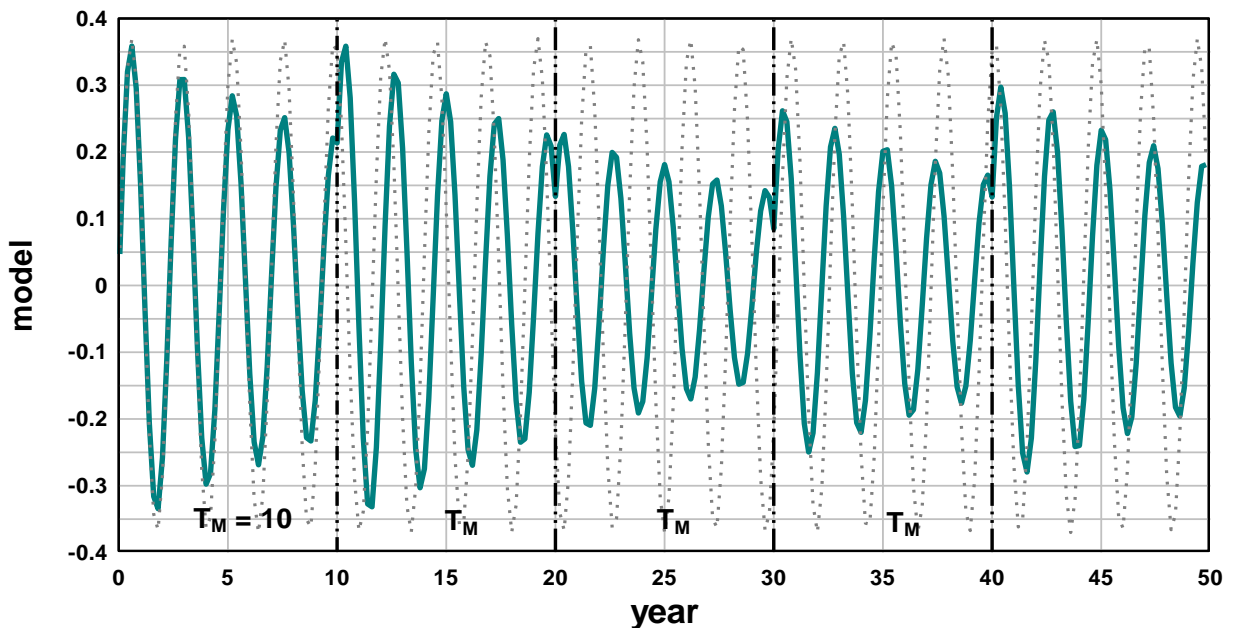

Figure 29. The summed responses of Figure 28 simulate the response of a simple harmonic oscillator to acceleration impulses at intervals of 10 years. The dotted curve illustrates the response to an impulse at year zero when the damping is zero i.e. a continuous un-damped SHM. Clearly the response to multiple periodic impulses results in a response of varying amplitude and phase, i.e. a quasi-periodic response with disruptions of the oscillation pattern evident at the times of the acceleration impulses.

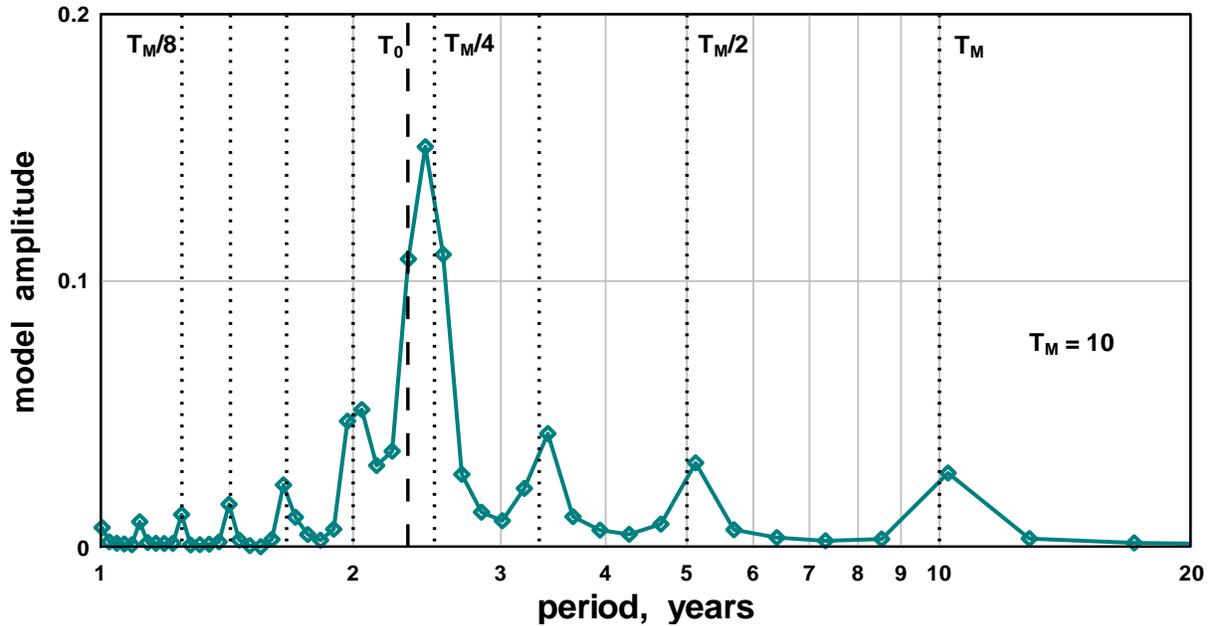

Figure 30. The spectrum of the model response indicates that: (1), the peak response is shifted from the natural period of the SHO, $T_0$, towards the period of the nearest $T_M/n$ harmonic; in this case from $T_0$ = 2.32 years towards $T_M/4$ = 10/4 = 2.5 years. (2), the response contains significant power at sidebands of period $T_M/n$.

The model response, Figure 29, and the periodogram of the model response, Figure 30, suggest the basic features of a QBO type oscillation responding to periodic acceleration impulses would be as follows:

(1) The time variation of a QBO would exhibit disruption of the oscillation pattern near the times of the acceleration impulses. The time variation of the model, Figure 29, shows features that are similar to anomalous features observed in the Earth QBO. One anomalous event observed in the QBO is the disruption of the oscillation between 2015 and 2016, Osprey et al (2016). That disruption resulted in the emergence of double peaks after 2015, c.f. Figure 25. Double peaks are evident in the model time variation shown in Figure 29 coincident with acceleration impulses at year 20 and year 30. Clearly, double peaks arise under this model when a new impulse response of the oscillator is not in phase with the sum of previous impulse responses, c.f. Figure 28. When double peaks occur the phase of the ongoing oscillation can suffer a large phase change. For example, after the disruption at year 20 in the model variation in Figure 29 the phase of the oscillation shifts by approximately $\pi$ radians before reverting to the previous phase condition at the next acceleration pulse, year 30. Exactly this type of $\pi$ phase shift was observed in the Earth QBO between 2015 and 2020, see Figure 1, Anstey et al (2021). Other, less obvious disruptions of the QBO cycle occurred around 1965 and 1990, c.f. Figure 25, and there is evidently a major disruption of the QBO around 1950. The times of QBO disruption correspond reasonably well with the times of occurrence of peaks in Sun acceleration c.f. Figure 5. The previous section noted disruptions of Jupiter's QQO also occurred at times, 1990 and 2007, close to times of Sun acceleration peaks.

(2) The model periodogram, Figure 30, indicates that the effect of impulsive acceleration on an oscillator is to shift power from the central oscillation period into sidebands at periods $T_M/n$, or equivalently into peaks at frequencies spaced at $1/T_M$ years$^{-1}$. This is broadly consistent with the observed spectrum of the QBO, Figure 27.

(3) The model time variation, Figure 29, and model periodogram, Figure 30, shows that the average periodicity of the oscillator is entrained by the impulse response to be close to the period of the nearest harmonic of the impulse period, in the case of the model, $T_M/4$. This is evident in Figure 30 where the natural variation of the model oscillator, at $T_0 = 2.3$ years, is shown as the dotted curve. Entrainment of the oscillator by the impulses shifts the average period of the oscillator towards the period, $T_M/4 = 2.5$ years, c.f. Figure 30. The model spectrum in Figure 30 helps to explain why the spectrum of the Earth QBO is dominated by a spectrum of periodicity associated with the harmonics, $T_A/n$, c.f. Figure 27. It also makes it easier to understand why the spectra of other climate oscillation indices such as the NAO and SOI would, due to entrainment, have dominant periodicities close to one or more of the harmonics, $T_A/n$. It is useful to note that this model may explain why all solar and climate periodicities, with the exception of those due to planet or solar rotation, will be quasi-periodic in the sense that the average period in one time interval may differ significantly from the average period in another time interval. For example the average model period in the interval 0 to 10 in Figure 29 is the natural period of the SHO, $T_0 = 2.3$ years, whereas the model period in the interval 15 to 25 is close to 2.5 years, or $T_M/4$, due to the occurrence of a double peak at 20 years.

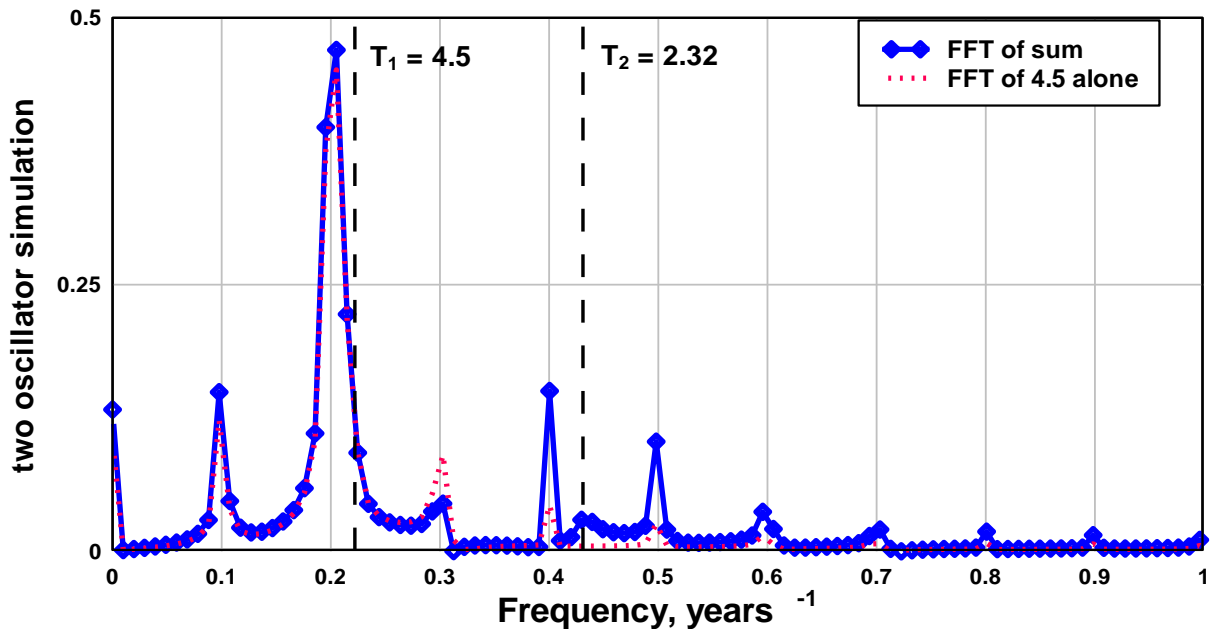

Figure 31. The model spectrum resulting when two simple harmonic oscillators, one at natural period 4.5 years the other at natural period 2.32 years, respond to periodic acceleration impulses at intervals of 10 years. Note that the natural periodicities are entrained to the next lower harmonic frequency of n/10; or in terms of period from 4.5 years to $T_M/2 = 5$ years and from 2.32 years to $T_M/4 = 2.5$ years.

(4) Figure 31 illustrates a fourth feature, i.e., if a system with multiple natural periods of oscillation is subjected to impulsive acceleration at intervals $T_M$ the natural periods will be entrained to shift towards the next harmonic period, $T_M/n$, above each natural period with the secondary sidebands to each contributing natural oscillator occurring at periods $T_M/n$ and the secondary sidebands either adding constructively or destructively, the latter effect illustrated at frequency 0.3 years$^{-1}$ in Figure 31.

It would be interesting to examine this model further to assess if there are circumstances where the entrainment of a natural oscillator could shift in period from $T_A/n$ to $T_A/(n+/-1)$ or even to $T_A/(n+/-3)$ as appears to have occurred with the QQO on Jupiter i.e. a shift from $T_A/7$ to $T_A/10$, 5.64 years to 3.95 years.

## 7. Summary of results

(1) The paper established that the mid – term periodicity of planet atmospheres is dominated by persistent periodicities with the periods occurring close periods in the series 39.5/n years where n = 2, 3, 4….

(2) It was shown that motion of the Sun relative to the barycentre results in strong acceleration impulses at intervals of $T_A$ = 39.5 years.

(3) As a result of (1) and (2) we conclude that the surfaces of the Sun, Earth, Jupiter and Saturn may respond to impulses of Sun acceleration with entrainment of the natural oscillations of the atmospheres to oscillations with periods close to $T_A/n$.

(4) For near monochromatic atmospheric oscillations like the QBO, disruption of the oscillation pattern is occasionally observed. Examples are, double peaks in the QBO. A very preliminary model based on the response of a simple harmonic oscillator to periodic impulses of acceleration appears to generate these types of anomalies.

Table 1 below compares the harmonics of $T_A$ with the observed periods of the strongest peaks in the mid frequency range of the various surface oscillations studied above. An entry with * indicates a value from the literature. For the near monochromatic QBO and the mesospheric quasi biennial oscillation (MQBO) that oscillates with period 60 months, close to $T_A/8$ years, Kumar (2021), only one period is included in Table 1. Similar tables that represent the correspondence between climate periodicity and some proposed driving periodicity, (however not, as for this Table, $T_A/n$), have been developed, (Velasco and Mendoza 2008, Pan et al 2020).

**Table 1. Compares the harmonic periods of $T_A$ with the periods of the strongest peaks of each example of atmospheric oscillation.**

| n | $T_A/n$ years | PDO | NAO | JUP | SAT | NEP | GSN | SOI | IOD | AO/AAO | QBO | MQBO |
|---|---|---|---|---|---|---|---|---|---|---|---|---|
| 2 | 19.8 | 18.6 | | | ~20* | | | | | | | |
| 3 | 13.2 | | 13.2 | | | 13* | | | | | | |
| 4 | 9.87 | 9.84 | | | | | 10.0 | | 10.7 | | | |
| 5 | 7.90 | | 7.79 | | | | 8.1 | | | | | |
| 6 | 6.58 | | | | | | | | | | | |
| 7 | 5.64 | 5.61 | | 5.7* | | | 5.7 | 5.71 | 5.3 | 5.68 | | |
| 8 | 4.94 | | 4.97 | | | | | | | | | 5* |
| 9 | 4.39 | | | | | | | | | | | |
| 10 | 3.95 | | | 3.9* | | | | | | | | |
| 11 | 3.59 | | | | | | | 3.53 | | 3.63 | | |
| 12 | 3.29 | | | | | | | | | | | |
| 13 | 3.04 | | | | | | | | 3.0 | | | |
| 14 | 2.82 | | | | | | | 2.85 | | 2.77 | | |
| 15 | 2.63 | | | | | | | | | | | |
| 16 | 2.47 | | | | | | | | | | | |
| 17 | 2.32 | | | | | | | | | | 2.34 | |

5. Disruptions in the QBO time variation occur at times near, 1950, 1965, 1990 and 2015 that correspond closely with times of impulsive Sun acceleration. The next significant disruption of atmospheric periodicity, likely to be observable in the QBO and in the QQO, should, on the evidence of this paper, occur around 2030, c.f. Figure 5.

6. The solar activity response at $T_A/6 = 6.58$ years is suppressed, c.f. Figure 10. A suppressed response of planet surface oscillation at $T_A/6$ is evident in Table 1 consistent with a cause/effect sequence from solar activity to climate oscillation.

## 8. Discussion and conclusion

The periodicity associated with atmospheric oscillations is complex. It ranges from near single period oscillations, for example the QBO, to complex, for example the NAO where numerous periodicities are evident. Where the oscillations occur in different regions of a planet, for example the North Atlantic in the case of the NAO and in the Pacific in the case of the SOI, the spectra of each oscillation can appear to be quite unrelated. For example, if the NAO and PDO spectra in Figure 1 were viewed without the reference lines the spectra would, at first sight, appear to be unrelated. However, when the complex and apparently unrelated spectra are compared with the harmonic series indicated by the reference lines in Figure 1 the probability that both oscillations are harmonically related to a single driving source of period 39.5 years becomes apparent. Other studies attempting to link periodicity in climate variability to harmonics of driving sources include Pan et al (2020) as discussed in the Introduction.

It is well known that the Fourier transform of a sequence of impulses at period T years is a set of impulses in the frequency domain separated by frequency 1/T years$^{-1}$ forming a harmonic series of frequencies given by n/T where n = 1 , 2, 3, .. etc.  If an atmosphere has a natural mode of oscillation or several modes of natural oscillation, then, if the atmosphere is subject to impulsive acceleration of period T, it is more likely to respond most strongly at one or more of its natural modes that are close in frequency to frequencies in the harmonic series n/T.  Further, the model developed in Section 6 suggests that natural modes of oscillation would be entrained towards average periodicity close to T/n. This would explain the observed correspondence in period of the peaks in spectra of different atmosphere oscillations to periods at 39.5/n years. So the question becomes – what is the physical mechanism by which a sequence of acceleration impulses experienced by the Sun that could influence atmospheric oscillations?

The Sun experiences impulsive accelerations and torques at time intervals of 39.5 years when it passes close to the barycentre (Jose 1965, Cionco 2008, Cionco and Abuin 2016) and Figure 5 of this paper. Also, strong responses in the solar QBO of flare activity are evident at times, for example 1990, as observed by Deng et al (2019), that we note are the same as the times of strong peaks in solar acceleration. Thus it is reasonable to advance the idea that the spectrum of periodicities at 39.5/n years observed in Sun acceleration, solar activity and atmospheric oscillations of the atmospheres of Earth and Jupiter are connected in some manner. The mechanism by which Sun acceleration stimulates solar activity and the mechanism by which solar activity influences climate variation are outside the scope of this paper. Various mechanisms of planet influence on solar activity have been proposed (Abreu et al 2012, Wolf and Patrone 2010, Scafetta 2012, Stefani et al 2021).  However, the nature of the mechanism, if it exists, remains uncertain, Charbonneau (2010).

**Appendix  A.  Illustrating the forward projection of climate variability.**

One reason for studying climate variables like the PDO, NAO and SOI is to try to understand the origin of the variability as in this paper. Another reason is to use knowledge about a climate variable to predict future variability. This is important as the climate in various regions is strongly dependent on the state, positive or negative, of these climate variables. Prediction is the major function of meteorology. The BOM, BOM (2023), predicts the ENSO index 6 months ahead. The IRI, IRI (2023), predicts the ENSO index eight months ahead.   Here we illustrate how knowledge of the spectral content of a climate variable, here the SOI, could be used to predict future climate and the caveats that apply. We emphasise that this is an illustration only. Figure A1A shows the monthly SOI between 1876 and 2023, a measure of the surface pressure difference between Tahiti and Darwin.  Large positive/negative values determine, to a large extent, the occurrence of la Nina/el Nino events and whether Australia experiences wet/dry conditions or, in the extreme, flood/drought. The full black line in Figure A1 is the sum of five narrowband components of the SOI centred on periods  $T_A/4$, 9.87 years, $T_A/7$, 5.64 yr, $T_A/11$, 3.59 yr, $T_A/14$, 2.82 yr and $T_A/17$, 2.32 years. The correlation coefficient between the SOI and the component sum is 0.64. How the components are found is illustrated in Figure A1B.

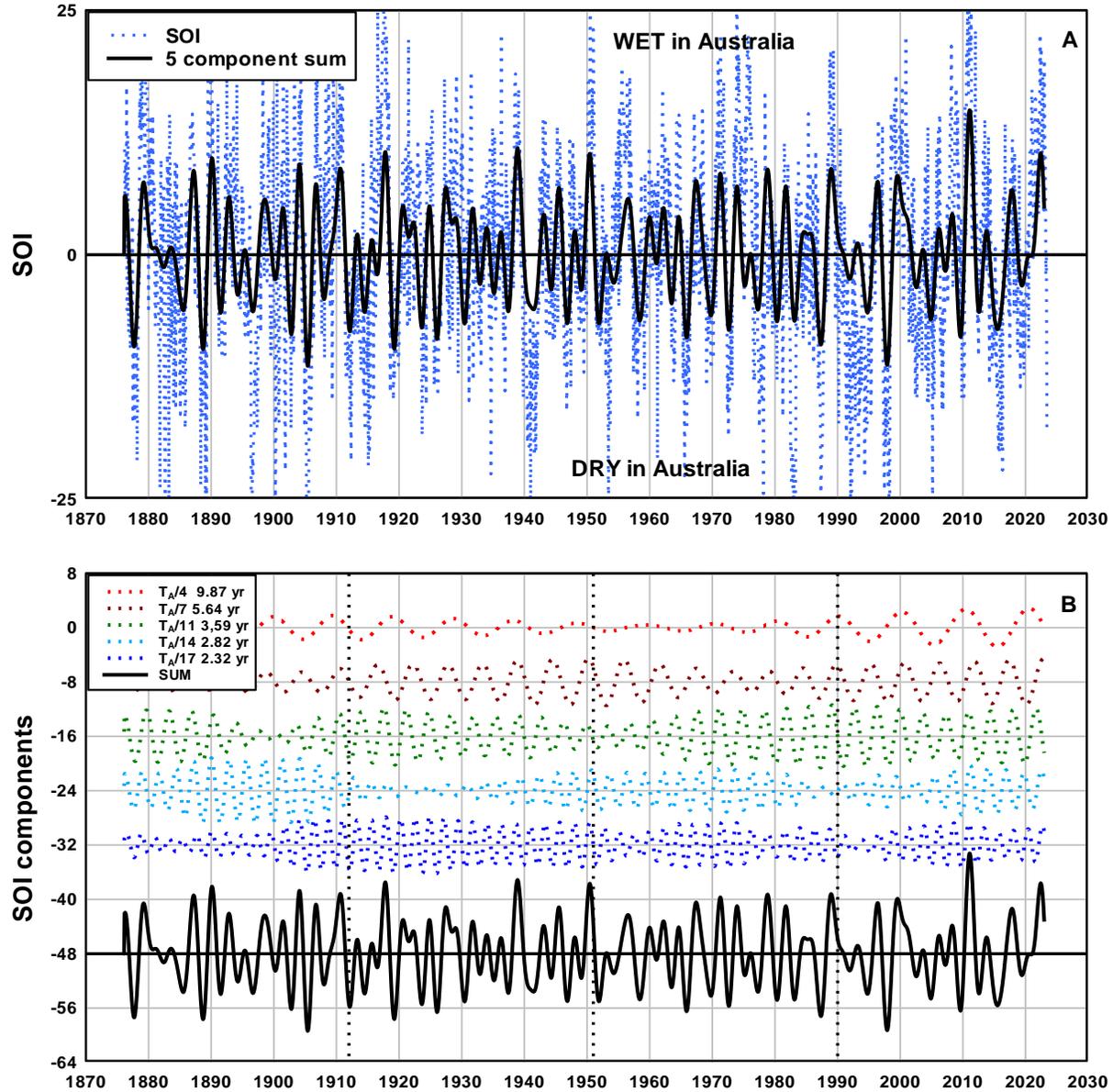

**Figure A1. (A) The monthly SOI index and the sum of five mid – frequency components. (B) The five mid frequency components of the SOI, displaced by eight SOI units for clarity and the sum of the five components. The reference lines are at the times when peaks in Sun acceleration occur.**

The mid frequency components are obtained by INF filtering the SOI with the centre period set to $T_A/n$ where n = 4, 7, 11, 14 and 17. The centre periods correspond to the periods of the five most prominent peaks in SOI spectrum, c.f. Figure 14. The bandwidth of the INF was set to 5% of each centre frequency. The sum of the five components is the full line curve in Figures A1A and A1B. It is evident from the comparison in Figure A1A that the five component sum is highly correlated with the SOI, i.e. the five component sum is representative of the medium time range, 2 – 10 year period, variation of the SOI. In Figure A1B it is evident that each component undergoes significant phase shifts at various times. For example, the $T_A/11$ component shifts at 1900, as discussed in section 4.12, the $T_A/14$ and $T_A/17$ components shift at around 1990. However, from 2010 onwards all five components appear to be in a stable phase condition that suggests the possibility of forward projecting from the 2010 – 2023 interval into the future as illustrated in Figure A2.

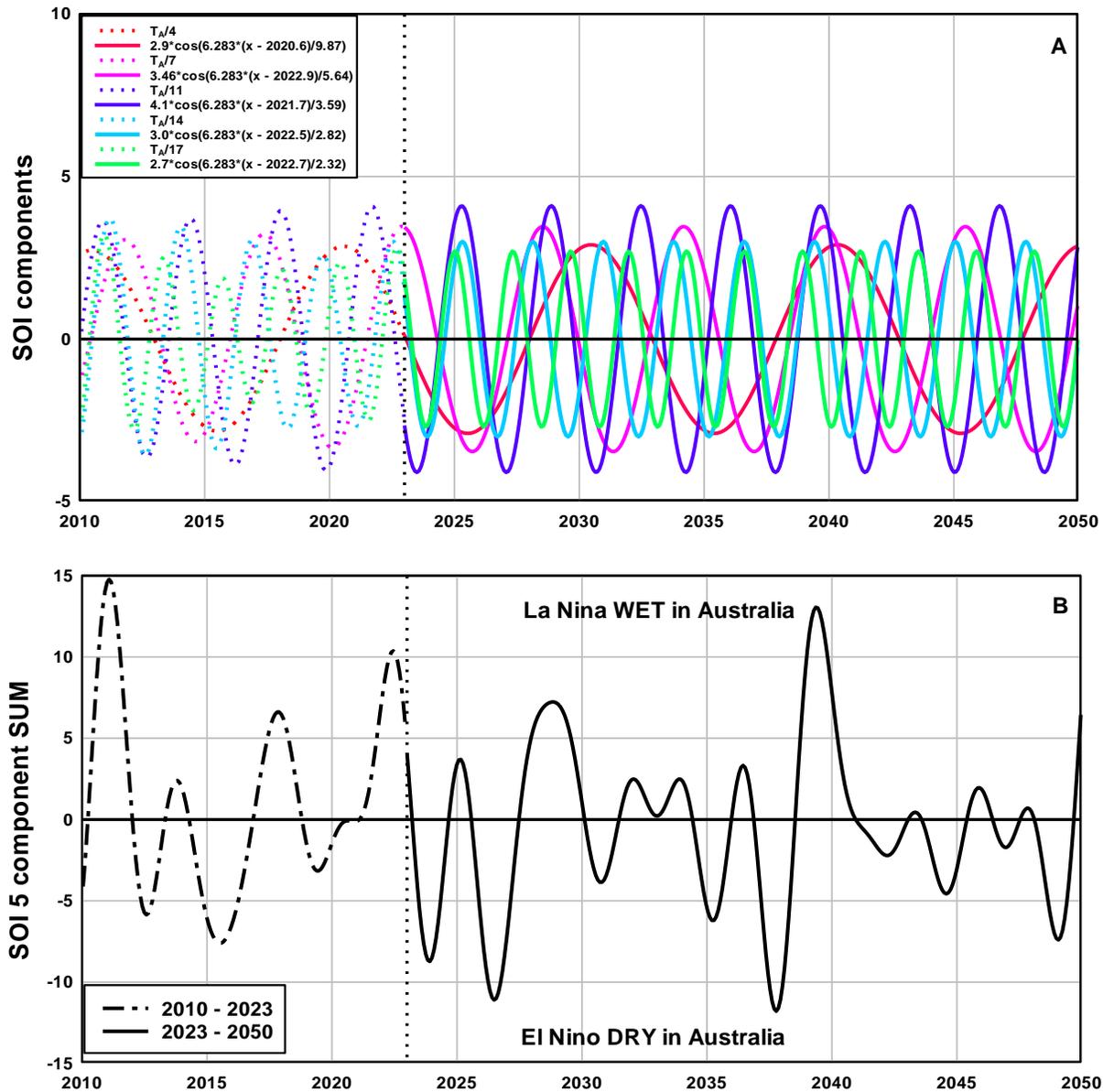

**Figure A2. (A) The five components of SOI in the 2010 – 2023 time interval (dotted curves) and the projected components (full line curves) 2023 - 2050. (B) The sum of the components 2010 – 2023 and the sum of the projected components 2023 – 2050.**

There are a range of methods for projecting climate variables, (Aprila et 2021, Abbot and Marohasy 2015, Barwain et al 2005, Schepen et al 2012, Nooteboom et al 2018), most using neural network theory. Here the projection is obtained by fitting sinusoids, of the form $A\cos(2\pi(t - XXXX)/T)$, to the components in the 2010 – 2023 interval and projecting the sinusoids forward to 2050. The full line in Figure A2B is the resulting projection of the SOI. The projection suggests a dry interval 2023 – 2028, a wet interval 2028 – 2030, average conditions, 2030 - 2037, a short dry spell, 2037- 2038, a wet spell 2037 - 2041 and average conditions 2041 - 2050. In the interval of expected greater accuracy, 2023 – 2030, as discussed below, the illustration projects two consecutive el Ninos followed by a la Nina.

The caveat applying to this projection illustration is that the accuracy of the projection depends on the projected components maintaining a stable phase condition. Previous work in this paper indicated that significant phase shifts in narrow band components are likely to occur at the time of the strong Sun acceleration impulses that occur at intervals of 39.5 years. The next strong acceleration impulse will occur at 2030 and it is likely that one or more of the SOI components used in Figure A2 will suffer a significant phase shift at, or soon after, 2030. Thus the expectation is that the projection 2023 - 2030 is the more accurate while the projection beyond 2030 is likely to be less accurate.

**References.**